\documentclass[traditabstract]{aa}
\usepackage[varg]{txfonts}

\usepackage{epsfig,graphicx,lscape,subfigure}
\usepackage{float,longtable,comment}
\usepackage{rotating,amssymb,amsmath}
\usepackage{natbib}

\newcommand{\kms}{{km\,s}$^{-1}$}
\newcommand{\teff}{$T_\mathrm{eff}$\,}
\newcommand{\logg}{$\log g$\,}

\newcommand{\Msun}{\,$\rm{M}_\odot$}
\newcommand{\Lsun}{\,$\rm{L}_\odot$}

\newcommand{\Mi}{\,$\rm{M}_{\rm{i}}$}

\newcommand{\MgtoH}{$\epsilon_{\rm{Mg}}$}
\newcommand{\NtoH}{$\epsilon_{\rm{N}}$}
\newcommand{\NtoHL}{$\epsilon_{\rm{0}}$}

\newcommand{\vsini}{$v_{\rm{e}} \sin i$}
\newcommand{\ve}{$v_{\rm{e}}$}
\newcommand{\vt}{$v_{\rm{t}}$}
\newcommand{\vi}{$v_{\rm{i}}$}
\newcommand{\vr}{$v_{\rm{r}}$}
\newcommand{\Dvr}{$\Delta v_{\rm{r}}$}

\DeclareRobustCommand{\rchi}{{\mathpalette\irchi\relax}}
\newcommand{\irchi}[2]{\raisebox{0.35ex}{$#1\chi$}}

\begin{document}

\title{The VLT-FLAMES Tarantula Survey\thanks{Based on observations at the European Southern Observatory in programmes 171.D0237, 073.D0234 and 182.D-0222} }
\subtitle{XXVIII. Nitrogen abundances for apparently single dwarf and giant B-type stars with small projected rotational velocities}

\author{P.L. Dufton\inst{\ref{i1}}, A. Thompson\inst{\ref{i1}}, P. A. Crowther\inst{\ref{i2}}, C. J. Evans\inst{\ref{i3}}, F.R.N. Schneider\inst{\ref{i4}}, A. de Koter\inst{\ref{i5}}, S. E. de Mink\inst{\ref{i5}},  R. Garland\inst{\ref{i1},\ref{i1a}}, N. Langer\inst{\ref{i6}}, D. J. Lennon\inst{\ref{i7}}, C. M. McEvoy\inst{\ref{i1},\ref{i1b}}, O.H. Ram\'irez-Agudelo\inst{\ref{i3}}, H. Sana\inst{\ref{i8}},  S. S\'imon D\'iaz\inst{\ref{i9},\ref{i10}}, W. D. Taylor\inst{\ref{i3}}, J. S. Vink\inst{\ref{i11}}}
% 1  Philip Dufton, Catherine McEvoy, Andrew Thompson, Ryan Garland
\institute{Astrophysics Research Centre, School of Mathematics and Physics, Queen's University Belfast, Belfast BT7 1NN, UK\label{i1}
% 2 Paul Crowther
    \and{Department of Physics and Astronomy, Hounsfield Road, University of Sheffield, Sheffield, S3 7RH, UK}\label{i2}
%  3 Chris Evans; William Taylor
	\and UK Astronomy Technology Centre, Royal Observatory Edinburgh, Blackford Hill, Edinburgh, EH9 3HJ, UK\label{i3}
%  4 Fabian Schneider
    \and Department of Physics, University of Oxford, Keble Road, Oxford OX1 3RH, United Kingdom\label{i4}	
%  5 Selma de Mink
    \and Anton Pannenkoek Institute for Astronomy, University of Amsterdam, NL-1090 GE Amsterdam, The Netherlands\label{i5}
% 1a Ryan Garland
        \and Sub-department of Atmospheric, Oceanic and Planetary Physics, Department of
Physics, University of Oxford, Oxford, OX1 3RH, UK \label{i1a}
% 6 Norbert Langer
         \and Argelander-Institut f\"{u}r Astronomie der Universit\"{a}t Bonn, Auf dem H\"{u}gel 71, 53121 Bonn, Germany\label{i6}
% 7 Danny Lennon
    \and European Space Astronomy Centre (ESAC), Camino bajo del Castillo, s/n Urbanizacion Villafranca del Castillo, Villanueva de la Ca\~{n}ada, E-28692 Madrid, Spain\label{i7}
% 1b Catherine McEvoy	
    \and King's College London, Graduate School, Waterloo Bridge Wing, Franklin Wilkins Building, 150 Stamford Street, London SE1 9NH, UK\label{i1b}
% 8 Hugues Sana
     \and Instituut voor Sterrenkunde, Universiteit Leuven, Celestijnenlaan 200 D, B-3001 Leuven, Belgium \label{i8}
% 9, 10 Sergio Simon Diax
   \and Instituto de Astrof\'isica de Canarias, E-38200 La Laguna, Tenerife, Spain\label{i9}
   \and Departamento de Astrof\'isica, Universidad de La Laguna, E-38205 La Laguna, Tenerife, Spain\label{i10}
% 11 Jorick Vink
   \and Armagh Observatory, College Hill, Armagh, BT61 9DG, Northern Ireland, UK\label{i11}
%        %\and {Space Telescope Science Institute, 3700 San Martin Drive, Baltimore, MD 21218, USA}
%% 5 Ian Howarth
%        \and{Department of Physics and Astronomy, University College London, Gower Street, London WC1E 6BT, UK}
%        \and{Department of Physics and Astronomy, Johns Hopkins University, 3400 North Charles Street, Baltimore, MD 21218, USA}
%     \and{Astronomical Institute `Anton Pannekoek', University of Amsterdam, Postbus 94249, 1090 GE, Amsterdam, The Netherlands}
%     \and{Instituto de Astrof\'isica de Canarias, E-38200 La Laguna, Tenerife, Spain}              
%        \and{Departamento de Astrof\'isica, Universidad de La Laguna, E-38205 La Laguna, Tenerife, Spain}
%     \and {Scottish Universities Physics Alliance, Institute for Astronomy, University of Edinburgh, Royal Observatory Edinburgh, Blackford Hill, Edinburgh, EH9 3HJ, UK}}
%\email{p.dufton@qub.ac.uk}}
}
\date{Received / accepted }

\abstract{Previous analyses of the spectra of OB-type stars in the Magellanic Clouds have identified targets with low projected rotational velocities and relatively high nitrogen abundances; the evolutionary status of these objects remains unclear. The VLT-FLAMES Tarantula Survey  obtained spectroscopy for over 800 early-type stars in 30 Doradus of which 434 stars were classified as B-type.  We have estimated atmospheric parameters and nitrogen abundances  using {\sc tlusty} model atmospheres for 54 B-type targets that appear to be single, have projected rotational velocities, \vsini\,$\leq$\,80 \kms and were not classified as supergiants. In addition, nitrogen abundances for 34 similar stars observed in a previous FLAMES survey of the Large Magellanic Cloud have been re-evaluated. 

For both samples, approximately 75-80\% of the targets have nitrogen enhancements of less than 0.3 dex, consistent with them having experienced only small amounts of mixing. However, stars with low projected rotational velocities, \vsini\,$\leq$\,40 \kms\ and significant nitrogen enrichments are found in both our samples and simulations imply that these cannot all be rapidly rotating objects observed near pole-on. For example,  adopting an enhancement threshold of 0.6 dex, we observed five and four stars in our VFTS and previous FLAMES survey samples, yet stellar evolution models with rotation predict only 1.25$\pm$1.11 and 0.26$\pm$0.51 based on our sample sizes and random stellar viewing inclinations. The excess of such objects  is estimated to be 20-30\% of all stars with current rotational velocities of less than 40\,\kms. This would correspond to $\sim$2-4\% of the total non-supergiant  single B-type sample. Given the relatively large nitrogen enhancement adopted, these estimates constitute lower limits for stars that appear inconsistent with current grids of stellar evolutionary models. Including targets with smaller nitrogen enhancements of greater than 0.2 dex implies larger percentages of targets that are inconsistent with current evolutionary models, viz.  $\sim$70\% of the stars with rotational velocities less than 40\,\kms and $\sim$6-8\% of the total single stellar population. We consider possible explanations of which the most promising would appear to be breaking due to magnetic fields or stellar mergers with subsequent magnetic braking.}

\authorrunning{P.L. Dufton et al.}
\titlerunning{B-type stellar nitrogen abundances in the Tarantula Nebula}

\maketitle
\nopagebreak[4]

%
%________________________________________________________________
\section{Introduction}                                         \label{s_intro}

The evolution of massive stars during their hydrogen core burning phase has been modelled for over sixty years with early studies by, for example, \citet{tay54, tay56, kus57, sch58, hen59, hoy60}. These initial models lead to a better understanding of, for example,  the observational Hertzsprung-Russell diagram for young clusters \citep{hen59} and the dynamical ages inferred for young associations \citep{hoy60}.  Subsequently there have been major advances in the physical assumptions adopted in such models, including the effects of stellar rotation \citep{mae87}, mass loss \citep[particularly important in more luminous stars and metal-rich environments;][]{chi86, pul08, mok07} and magnetic fields \citep{don09,pet15}. Recent reviews of these developments include \citet{mae09} and \citet{lan12}.

Unlike late-type stars, where stellar rotation velocities are generally small \citep[see, for example,][and references therein]{gra05, gra16, gra17}, rotation is an important phenomenon affecting both the observational and theoretical understanding of early-type stars.  Important observational studies of early-type stellar rotation in our Galaxy include \citet{sle49}, \citet{con77}, \citet{pen96}, \citet{how97}, \citet{abt02}, \citet{hua10} and \citet{sim14}. There have also been extensive investigations  for the Magellanic Clouds using ESO large programmes, viz. \citet[][and references therein]{mar06, mart07}, \citet{eva05, eva06} and \citet{hun08a}. More recently as part of the VLT-Flames Tarantula Survey \citep{eva11}, rotation in both apparently single \citep{ram13, duf12} and binary \citep{ram15} early-type stars has been investigated. These studies all show that early-type stars cover the whole range of rotational velocities up to the critical velocity at which the centrifugal force balances the stellar gravity at the equator \citep{str31, tow04}.

These observational studies have led to rotation being incorporated into stellar evolutionary models \citep[see, for example,][]{mae87, heg00b, hir04, fri10}, leading to large grids of rotating evolutionary models for different metallicity regimes \citep{bro11a, eka12, geo13, geo13a}. Such models have resulted in a better understanding of the evolution of massive stars, including main-sequence life times \citep[see, for instance][]{mey00, bro11a, eka12} and the possibility of chemically homogeneous evolution \citep{mae80, mae12, sze15}, that in low-metallicity environments could lead to gamma-ray bursts \citep{yoo05, yoo06, woo06}. However there remain outstanding issues in the evolution of massive single and binary stars \citep{mey17}. For example, nitrogen enhanced early-type stars with low projected rotational velocities that may not be the result of rotational mixing have been identified by \citet{hun08b}, \citet{riv12}, and \citet{gri17}. The nature of these stars have discussed previously by, for example, \citet{bro11b}, \citet{mae14}, and \citet{aer14a}. Definitive conclusions have been hampered both by observational and theoretical uncertainties and by the possibility of other evolutionary scenarios \citep[see][for a recent detailed discussion]{gri17}. 

Here we discuss the apparently single B-type stars with low projected rotational velocity found in the VLT-Flames Tarantula Survey \citep[hereafter VFTS]{eva11}. These complement the analysis of the corresponding binary stars from the same survey by \citet{gar17}. We also reassess the results of \citet{hun07, hun08b} and \citet{tru08} from a previous spectroscopic survey \citep{eva06}. In particular we investigate whether there is an excess of nitrogen enhanced stars with low projected velocities and attempt to quantify any such excess. We then use these results to constrain the evolutionary pathways that may have produced these stars.

\section{Observations}\label{s_obs}

The VFTS spectroscopy was obtained using the MEDUSA and ARGOS modes of the FLAMES instrument \citep{pas02} on the ESO Very Large Telescope. The former uses fibres to simultaneously `feed'  the light from over 130 stellar targets or sky positions to the Giraffe spectrograph. Nine fibre configurations (designated fields~`A' to `I' with near identical field centres) were observed in the 30~Doradus region, sampling the different clusters and  the local field population.  Spectroscopy of more than 800 stellar targets was obtained and approximately half of these were subsequently spectrally classified by \citet{eva15} as B-type.   The young massive cluster at the core of 30~Doradus, R136, was too densely populated for the MEDUSA fibres, and was therefore observed with the ARGUS integral field unit (with the core remaining unresolved even in these observations).  Details of target selection, observations, and initial data reduction have been given in \citet{eva11},  where target co-ordinates are also provided.

The analysis presented here employed the FLAMES--MEDUSA observations that were obtained with two of the standard Giraffe settings, viz.  LR02 (wavelength range from 3960 to 4564\,\AA\ at a spectral resolving power of R$\sim$7\,000) and LR03 (4499-5071\,\AA, R$\sim$8\,500). The targets identified as B-type (excluding a small number with low quality spectra) have been analysed using a cross-correlation technique to identify radial velocity variables \citep{dun15}.  For approximately 300 stars, which showed no evidence of significant radial velocity variations, projected rotational velocities, \vsini, have been estimated by \citet{duf12} and \citet{gar17}. 

In principle, a quantitative analysis of all these apparently single stars, including those with large projected rotational velocities, would have been possible. However, the moderate signal-to-noise ratios (S/N) of our spectroscopy would lead to the atmospheric parameters (especially the effective temperature and microturbulence -- see Sect. \ref{s_teff} and \ref{s_vt}) being poorly constrained for the broader lined targets.  In turn this would lead to nitrogen abundance estimates that had little diagnostic value. Hence for the purposes of this paper we have limited our sample to the subset with relatively small projected rotational velocities. This is consistent with the approach adopted by \citet{gar17} in their analysis of the corresponding narrow lined B-type VFTS binaries.

Hence our targets were those classified as B-type with a luminosity class III to V,  which showed no evidence of significant radial velocity variations and had an estimated \vsini\,$\le$80 \kms; the B-type supergiants (luminosity classes I to II) have been discussed previously by \citet{mce14}. Stars with an uncertain O9 or B0 classification have not been included. Additionally the spectrum of VFTS\,469 had been identified in Paper I as suffering some cross contamination from a brighter O-type star in an adjacent fibre. Another of our targets, VFTS\,167,  may also suffer such contamination as the \ion{He}{ii} line at 4686\,\AA\ shows broad emission. We retain these two stars in our sample but discuss the reliability of their analyses in Sect.\ \ref{s_GC}. Some of our apparently single stellar sample will almost certainly have a companion. In particular, as discussed by \citet{dun15}, low mass companions or binaries in wide orbits may not have been identified. 

Table~\ref{t_Targets} summarises the spectral types \citep{eva15} and typical S/N (rounded to the nearest multiple of five) in the LR02 spectral region. The latter were estimated from the wavelength region, 4200-4250\,\AA, which should not contain strong absorption lines. However particularly for the higher estimates, these should be considered as lower limits as they could be affected by weak lines. S/N  for the LR03 region were normally similar or slightly smaller \citep[see, for example,][for a comparison of the ratios in the two spectral region]{mce14}. Also listed are the estimated projected rotational velocity,  \vsini\ \citep{duf12, gar17}, the mean radial velocity, \vr\ \citep{eva15}, and the range, \Dvr, in radial velocity estimates \citep{dun15}. The moderate spectral resolving power led to estimates of the projected rotational velocity for stars with \vsini\,$\leq$ 40 \kms\ being poorly constrained and hence approximately half the sample have been  assigned to a bin with $0\le$\,\vsini\,$\le40$\ \kms.

Luminosities have also been estimated for our targets using a similar methodology to that adopted by \citet{gar17}. Interstellar extinctions were estimated from the observed $B-V$ colours provided in \citet{eva11}, together with the intrinsic colour--spectral-type calibration of \citet{weg94} and $R_{V} = 3.5$ \citep{dor13}. Bolometric corrections were obtained from LMC-metallicity {\sc tlusty} models \citep{lan07} and together with an adopted LMC distance modulus of 18{\fm}49 \citep{pie13},  the stellar luminosity were estimated. Table \ref{t_Targets} lists these estimates (hereafter designated as `observed luminosities') apart from that for VFTS\,835 for which no reliable photometry was available. The main sources of uncertainty in these luminosity estimates will arise from the bolometric correction and the extinction. We estimate that these will typically contribute an error of $\pm 0^{\mathrm{m}}$\!\!.5 in the absolute magnitude corresponding to $\pm$0.2 dex in the luminosity. We note these error estimates are consistent with the luminosities deduced from evolutionary models discussed in Sect.\ \ref{d_bonnsai}.

For both the LR02 and LR03 spectroscopy \footnote{The DR2.2 pipeline reduction was adopted and the spectra are available at http://www.roe.ac.uk/$\sim$cje/tarantula/spectra}, all useable exposures were combined using either a median or weighted $\sigma$-clipping algorithm. The final spectra from the different methods were  normally indistinguishable.  The full wavelength range for each wavelength setting could usually be normalised using a single, low-order polynomial. However, for some features (e.g. the Balmer series), the combined spectra around individual (or groups of) lines were normalised.

In the model atmosphere analysis, equivalent widths were generally used for the narrower metal lines, with profiles being considered for the broader  \ion{H}{i} and \ion{He}{ii} lines. For the former, Gaussian profiles (and a low order polynomial to represent the continuum) were fitted to the observed spectra leading to formal errors in the equivalent width estimates of typically 10\%. The fits were generally convincing which is consistent with the intrinsic and instrumental broadening being major contributors to these narrow lined spectra. Tests using rotationally broadened profiles yielded equivalent width estimates in good agreement with those using Gaussian profiles -- differences were normally less than 10\% but the fits were generally less convincing.

\begin{figure}
\epsfig{file=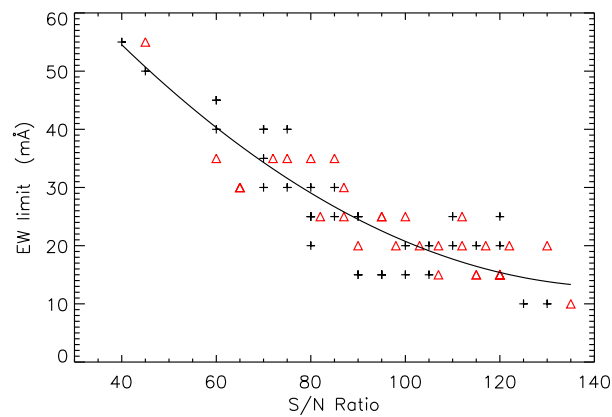,width=9cm, angle=0} 
\caption{Estimated equivalent width limits (in m\AA) plotted against the S/N of the VFTS spectra. Crosses and triangles represent stars with \vsini\,$\leq$\,40 \kms\ and 40\,$<$\,\vsini\,$\leq$\,80 \kms\ respectively. A quadratic linear least squares fit for all the estimates is also shown.}
\label{ew_sn}
\end{figure}

For some spectra, lines due to important diagnostic species (e.g. \ion{Si}{ii}, \ion{Si}{iv} and \ion{N}{ii}) could not be observed; in these cases we have set upper limits on the equivalent widths. Our approach was to measure the equivalent width of the weakest metal line or lines that were observable in the VFTS spectra of narrow lined stars. These were then rounded to the nearest 5\,m\AA\  and plotted against the S/N of the spectra as shown in Fig. \ref{ew_sn}. The equivalent widths decrease with increasing S/N as would be expected and there also seems to be no significant difference for stars with 0\,$<$\,\vsini\,$\leq$40 \kms\ and those with 40\,$<$\,\vsini\,$\leq$\,80 \kms. The upper limit for an equivalent width at a given S/N probably lies at or near the lower envelope of this plot. However we have taken a more cautious  approach by fitting a low order polynomial to these results and using this to estimate upper limits to equivalent  widths of unseen lines. We emphasise that this will lead to conservative estimates both for the upper limits of equivalent widths and for the corresponding element abundances.

The VFTS data have been supplemented by those from a previous FLAMES/VLT survey \citep[][hereafter the FLAMES-I survey]{eva05}. Spectral classification by \citet{eva06} identified 49 B-type stars that also had \vsini\,$\le$\,80 \kms\ \citep{hun08a}. Fourteen of these were characterised as binaries \citep{eva06}, either SB2 (on the basis of their spectra) or SB1 (on the basis of radial velocity variations between epochs). These targets were discussed by \citet{gar17} and will not be considered further. For the apparently single stars,  atmospheric parameters previously estimated from this FLAMES-I spectroscopy  \citep{hun07, hun08b, hun09a, tru07} have been adopted.  However we have re-reduced the spectroscopy for the wavelength region incorporating the  \ion{N}{ii} line at 3995\,\AA, using the methods outlined above. This should provide a better data product than the simple co-adding of exposures that was under taken in the original analyses. Comparison of spectra from the two reductions confirmed this (although in many cases the improvement was small). 
% 49 stars - 36 single/13 binary

\section{Binarity criteria} \label{s_bin}
\citet{dun15} adopted two criteria for the identification of binary systems\footnote{Some of these systems could contain more than two stars.}. Firstly the difference between two estimates of the radial velocity had to be greater than four times the estimated uncertainty and secondly this difference had to be greater than 16 \kms; this threshold was chosen so as to limit false positives due to for example pulsations that are likely to be significant particularly in supergiants \citep{sim10, tay14, sim17a}. In Table \ref{t_Targets}, the ranges (\Dvr) of radial velocities of six targets fulfil the latter (but not the former) criteria. As discussed by \citet{dun15} a significant fraction of binaries may not have been identified and these six targets could therefore be binaries. Four of these targets (VFTS\,010, 625, 712, 772) were not analysed due to the relatively low S/N of their spectroscopy (see Sect. \ref{s_parameters}), which probably contributed their relatively large values of \Dvr.

For the other two targets, we have re-evaluated the spectroscopy as follow:
\\
\\
\noindent
{\bf VFTS\,313:} \citet{dun15} obtained radial velocity estimates only from the \ion{He}{i} diffuse lines at 4026 and 4387\,\AA, which may have contributed to the relatively large value of \Dvr\ $\sim$ 18 \kms. There was considerable scatter between estimates found for exposures obtained at a given epoch, especially the first epoch when six consecutive exposures were available. Given the high cadence of these observations, these variations are unlikely to be real and probably reflect uncertainties in the individual  measurements. The best observed intrinsically narrow feature in the LR02 spectra was the \ion{Si}{iv} and \ion{O}{ii} close blend (separation of 0.60\,\AA) at 4089\,\AA. Radial velocity estimates were possible for eleven out the twelve LR02 exposures (the other exposure was affected by a cosmic ray event) from fitting a Gaussian profile to this blend. The estimates ranged from 270-282\,\kms\ with a mean value of 275.1$\pm$4.4 \kms\ (assuming that the \ion{Si}{iv} feature dominates the blend). 
\\
\\
\noindent
{\bf VFTS\,835:} \citet{dun15} obtained radial velocity estimates only from  the \ion{He}{i} diffuse lines at 4026, 4387 and 4471\,\AA, which may again have contributed to the relatively large value of \Dvr\ $\sim$ 18 \kms. As for VFTS\,313, there was again considerable scatter between estimates found for exposures within a given epoch. Unfortunately the metal lines were broadened by stellar rotation, which precluded their use for estimating reliable radial velocities in the single exposures, which had relatively low S/N.
~\\

Hence we find no convincing evidence for binarity in either of these targets and have retained them in our apparently single star sample.

\begin{table}
 \caption{VFTS B-type targets with a luminosity class V to III \citep{eva15}, which show no evidence of significant radial velocity variations \citep{dun15, eva15} and have a \vsini\,$\le$\,80 \kms\ \citep{duf12,gar17}. The mean radial velocities, \vr\ are from \citet{eva15} and the ranges, \Dvr, in radial velocity estimates are from \citet{dun15} with all velocities having units of \kms.  Further detail on the targets including their co-ordinates can be found in \citet{eva11}. The estimates of the S/N for the LR02 region have been rounded to the nearest multiple of five, whilst luminosities, L, were estimated as discussed in Sect. \ref{s_obs}.}\label{t_Targets}
\begin{tabular}{llrrrrr}
\hline\hline
VFTS & ST & S/N & \vsini & \vr~~  & \Dvr & $\log$ L/\Lsun\\
\hline

010         &     B2 V                 &     40      &     $\le$40          &     287  & 25 & 3.87 \\
024         &     B0.2 III-II          &     100    &           58            &     286  & 10 & 4.80 \\
029         &     B1 V                 &     80      &     $\le$40          &     286  & 10 & 3.95\\
044         &     B2 V                 &     80      &          65             &     288  &   7 & 3.79\\
050         &     B0 V                 &     60      &     $\le$40          &     283   &  3  & 4.35\\
052         &     B0.2 III-II          &     145    &          48             &     280   &  4 & 4.77 \\
053         &     B1 III                &     125    &     $\le$40           &     276  &  3  & 4.56\\
075         &     B1 V                 &     65      &          70             &     297   &  3  & 3.94\\
095         &     B0.2 V              &     100    &     $\le$40          &     294   &  4  & 4.36\\
111          &     B2 III                &     215    &          80             &     272   &  2 & 4.56 \\
119          &     B0.7 V              &     60      &     $\le$40          &     276   &  11 & 4.27\\
121          &     B1 IV                &     115    &     $\le$40          &     284   &  3   & 4.31\\
124          &     B2.5 III             &     75      &     $\le$40          &     289   &  6  & 4.04 \\
126          &     B1 V                 &     80      &     $\le$40          &     254   &  3  & 3.86\\
152          &     B2 IIIe              &     130    &          47             &     274   &  4  & 4.65\\
167          &     B1 V                 &      110   &     $\le$40          &     286   &  14 & 4.26\\
170          &     B1 IV                &     120    &     $\le$40          &     279   &  1  & 4.41\\
183          &     B0 IV-III            &     70      &     $\le$40          &     256   & 2  & 4.66\\
202          &     B2 V                 &     120    &          49             &     277   & 7  & 4.25\\
209          &     B1 V                 &     80      &     $\le$40          &     273   & 14 & 4.06 \\
214          &     B0 IV-III            &     80      &     $\le$40          &     276   & 4   & 4.78\\
237          &     B1-1.5 V-IVe     &     110    &          79             &     288    & 10 &  4.35\\
241          &     B0 V-IV             &     75      &          69             &     268    & 16 & 4.14\\
242          &     B0 V-IV             &     60      &     $\le$40          &     216    & 8  & 4.17\\
273          &     B2.5 V              &     70      &     $\le$40          &     254    & 4 & 3.78\\
284          &     B1 V                 &     80      &     $\le$40         &     273    & 15 & 3.91 \\
297          &     B1.5 V              &     80      &          47            &     252    & 9 & 4.01\\
308          &     B2 V                 &     160    &          74            &     263    & 3  & 4.17\\
313          &     B0 V-IV             &     110    &          56            &     270    & 18  & 4.41\\
331          &     B1.5 V              &     100    &          64            &     279    & 3  & 3.84\\
347          &     B0 V                 &     95      &     $\le$40         &     268    & 4  & 4.19\\
353          &     B2 V-III             &     90      &          63            &     292    & 7 & 4.02\\
363          &    B0.2 III-II           &     225    &          50            &     264    & 7 & 4.64\\
384          &     B0 V-III             &     65      &          46            &     266    & 6 & 4.79\\
469          &     B0 V                 &    100     &     $\le$40         &     276    & 3 & 4.41 \\
478          &     B0.7 V-III          &     85      &          64            &     273    & 13 & 4.41\\
504          &     B0.7 V              &     95      &          60            &     251    &13  & 5.06\\
540          &     B0 V                 &     85      &          54            &     247    & 12 & 4.55 \\
553          &     B1 V                 &     85      &          50            &     277    & 7  & 3.96\\
572          &     B1 V                 &     115     &          68           &     249    & 12 & 4.33\\
593          &     B2.5 V              &     70      &     $\le$40         &     292    & 7 & 3.82\\
616          &     B0.5: V             &     100    &     $\le$40         &     227    & 8& 4.29 \\
623          &     B0.2 V              &     110    &     $\le$40         &     269    & 4 & 4.57\\
625          &     B1.5 V              &     65      &          64            &     310   & 41& 4.00 \\
650          &     B1.5 V              &     60      &     $\le$40         &     287   & 6 & 3.68\\
666          &     B0.5 V              &     95      &     $\le$40         &     255   & 3 & 4.07\\
668          &     B0.7 V              &     90      &     $\le$40         &     270   & 5 & 4.21\\
673          &     B1 V                 &     80      &     $\le$40         &     278  & 2 & 3.92\\
692          &     B0.2 V              &     95      &     $\le$40         &     271   & 3 & 4.42\\
707          &     B0.5 V              &     145    &     $\le$40         &     275   & 10  & 4.61\\
712          &     B1 V                 &     120    &          67            &     255   & 17 & 4.52\\
725          &     B0.7 III             &     120    &     $\le$40         &     286   &  5 & 4.39 \\
\hline
\end{tabular}
\end{table}

\setcounter{table}{0}
\begin{table}
  \caption{(continued.)}
\begin{tabular}{llrrrrr}
\hline\hline
VFTS & ST & S/N & \vsini & \vr~~  & \Dvr & $\log$ L/\Lsun  \\
&&& \multicolumn{3}{c}{\kms} \\
\hline
727          &     B3 III                &     70      &          76            &     272   & 7& 3.84 \\
740          &     B0.7 III             &     45      &     $\le$40         &     268   &  13 & 4.48\\
748          &     B0.7 V             &     85      &          62            &     261    & 11 & 4.23\\
772          &     B3-5 V-III         &     70      &          61            &     254    & 17 & 3.41\\
801          &     B1.5 V              &     75      &     $\le$40         &     278   & 7 & 4.13\\
835          &     B1 Ve               &     80      &          47            &     253   &18 & - \\
851          &     B2 III                &     120    &     $\le$40          &     229  & 2 & 3.97 \\
860          &     B1.5 V             &     75      &           60            &     253   & 3 & 4.12\\
864          &     B1.5 V             &     70      &     $\le$40         &     278  &  13 & 3.94\\%SNR from PLD  
868          &     B2 V                &     60      &     $\le$40         &     262   & 8& 4.25 \\
872          &     B0 V-IV            &     100    &          77           &     287    & 6 & 4.30\\
879          &     B3 V-III            &      40     &      $\le$40        &     276    & 16 & 3.77\\%SNR from PLD 
881          &     B0.5 III            &     80      &     $\le$40          &     283   & 3 & 4.28\\
885          &     B1.5 V             &     95     &          69             &     264    & 9 & 4.26\\
886          &     B1 V                &     65     &     $\le$40          &     258   & 14 & 3.89 \\
\hline
\end{tabular}
\end{table}
% 67 targets
% 37 <40
% 30 40-80

\section{Atmospheric parameters} \label{s_parameters}

\subsection{Methodology}\label{s_meth}

We have employed model-atmosphere grids calculated with the {\sc tlusty} and {\sc synspec} codes \citep{hub88, hub95, hub98, lan07}.  They cover a range of effective temperature, 10\,000K $\leq$\teff$\leq$35\,000K in steps of typically 1\,500K.  Logarithmic gravities (in cm s$^{-2}$)  range from 4.5 dex down to the Eddington limit in steps of 0.25 dex, and microturbulences are from 0-30 \kms\ in steps of 5 \kms. As discussed in  \citet{rya03} and \citet{duf05}, equivalent widths and line profiles interpolated within these grids are in good agreement with those calculated explicitly at the relevant atmospheric parameters.

These non-LTE codes adopt the `classical' stationary model atmosphere assumptions, that is plane-parallel geometry, hydrostatic equilibrium, and the optical spectrum is unaffected by winds. As the targets considered here have luminosity classes V to III, such an approach should provide reliable results. The grids assumed a normal helium to hydrogen ratio (0.1 by number of atoms) and the validity of this is discussed in Sect.\  \ref{s_GC}. Grids have been calculated for a range of metallicities with that for an LMC metallicity being used here. As discussed by \citet{duf05}, the atmospheric structure depends principally on the iron abundance with a value of 7.2 dex having been adopted for the LMC; this is consistent with estimates of the {\em current} iron abundance for this galaxy \citep[see, for example,][]{car08, pal15, lem17}. However we note that, although tests using the Galactic (7.5 dex) or SMC (6.9 dex) grids led to small changes in the atmospheric parameters (primarily to compensate for the different amounts of line blanketing), changes in  element abundances estimates were typically less than 0.05 dex.  Further information on the {\sc tlusty} grids are provided in \citet{rya03} and \citet{duf05} 

Thirteen of the targets listed in Table \ref{t_Targets} have not been analysed  (VFTS\,010, 044, 152,  504, 593, 625, 712, 727, 772, 801, 879, 885,  886).
% 5 <40; 8 40-80
%10 stars having a S/N$\le$80
Most of these stars had spectroscopy with relatively low S/N, whilst several also suffered  from strong contamination by nebular emission. The former precluded the estimation of the effective temperature from the silicon ionisation equilibrium, whilst the latter impacted on the estimation of the surface gravity. In turn this led to unreliable atmospheric parameters, and hence any estimates of (or limits on) the nitrogen abundances would have had limited physical significance.

The four characteristic parameters of a static stellar atmosphere, (effective temperature, surface gravity, microturbulence and metal abundances) are inter-dependant and so an iterative process was used to estimate these values.

\subsection{Effective Temperature} \label{s_teff}

Effective temperatures (\teff) were principally determined from the silicon ionisation balance. For the hotter targets, the \ion{Si}{iii}  (4552, 4567, 4574\,\AA) and \ion{Si}{iv}  (4089, 4116\,\AA)  spectra were used while for the cooler targets, the \ion{Si}{iii} and \ion{Si}{ii} (4128, 4130\,\AA) spectra were considered. The estimates of the microturbulence from the absolute silicon abundance (method 2 in Sect. \ref{s_vt}) were adopted.  In some cases it was not possible to measure the strength of either the  \ion{Si}{ii} or \ion{Si}{iv} spectrum. For these targets, upper limits were set on their equivalent widths, allowing limits to the effective temperatures to be estimated, \citep[see, for example,][ for more details]{hun07}. These normally yielded a range of  effective temperatures of less than 2000K, with the midpoint being adopted.

For the hotter stars, independent estimates were obtained from the \ion{He}{ii} spectrum (at 4541\AA\ and 4686\AA). There was good  agreement between the  estimates obtained using the two methods, with differences typically being 500-1\,000 K as can be seen from Table \ref{t_Atm_1}. A conservative stochastic uncertainty of $\pm$1\,000 K has been adopted for the adopted effective temperature estimates listed in Table \ref{t_Atm_2}.

% 56 stars analysed
\begin{table}

\caption{Estimates of the atmospheric parameters for the VFTS sample. Effective temperatures (\teff) are from the silicon ionisation balance (Si) or \ion{He}{ii} profiles -- those constrained by the absence of the \ion{Si}{ii} and \ion{Si}{iv} spectra are marked with an asterisk. Microturbulences (\vt) are from the relative strengths of the \ion{Si}{iii} triplet (method 1) or the absolute silicon abundance (method 2). Stars for which no solution could be found for the micrtoturbulence are identified by daggers.}\label{t_Atm_1}
\begin{tabular}{llllccc}
\hline\hline
\vspace*{-0.25cm}\\
VFTS  & \multicolumn{3}{c}{\teff (K)}  & \logg & \multicolumn{2}{c}{\vt\ (\kms)}  \\

& ~~~~Si & 4686\AA & 4541\AA &  (cm s$^{-2}$) &   (1) & (2)  \\

\hline
024          &     27500            &     27000             &     -      &     3.3     &   9   & \!\!\!10    \\
029          &     27000            &     27000     &     -              &     4.1     &  0   &  4     \\
050          &     30500            &     30500     &     30000     &     4.0     &   0  &   1    \\
052          &     27500            &     27000     &     -              &     3.4     &    \!\!\!11 &  9    \\
053          &     24500            &     -              &     -              &     3.5     &   0 &  4    \\
075          &     23500$^*$     &     -              &     -              &     3.8     &   0  &   6     \\
095          &     31000            &     30000     &     -              &     4.4     &   0   &     \,\,\,0$^{\dagger}$ \\
111          &     22000$^*$     &     -              &     -              &     3.3     &   -   &     \,\,\,0$^{\dagger}$ \\
119          &     27500            &     27000     &     -  & 4.2    &  \,\,\,0$^{\dagger}$ & \,\,\,0$^{\dagger}$ \\
121          &     24000            &     -              &     -              &     3.6     &   1  &     0     \\
124          &     20000            &     -              &     -              &     3.4     &   -  &     1     \\
126          &     23500$^*$     &     -              &     -              &     3.8     &   0  &     2     \\
167          &     27000            &     -              &     -              &     4.0     &   0  &     0     \\
170          &     23000            &     -              &     -              &     3.5     &   0  &     0  \\
183          &     29000            &     28500     &     28500     &     3.5     &   5  &     4     \\
202          &     23500$^*$     &     -              &     -              &     3.9    &   1   &     0     \\
209          &     27500            &     26500     &     -              &     4.3     &   0  &     0      \\
214          &     29000            &     29500     &     29500     &     3.6     &   3  &     3     \\
237          &     23000            &     -              &     -              &     3.6     &   0  &     2     \\
241          &     -                     &     30500     &     30500     &     4.0     &   -   &     0     \\
242          &     29500            &     29000     &     29000     &     3.9     &   0  &     0 \\
273          &     19000            &     -              &     -              &     3.8     &   -   &     5     \\
284          &     23500$^*$     &     -              &     -              &     3.9     &   1  &     0     \\
297          &     23500$^*$     &     -              &     -              &     3.6     &   2   &     \,\,\,0$^{\dagger}$ \\
308          &     23500$^*$     &     -              &     -              &     3.8     &   3   &     \,\,\,0$^{\dagger}$ \\
313          &     27500            &     30000     &     29000      &     3.8     &   0  &     \,\,\,0$^{\dagger}$ \\
331          &     24000$^*$     &     -              &     -              &     3.9     &   4   &     2     \\
347          &     30500            &     29500     &     30000      &     4.0     &   0  &     0     \\
353          &     23000$^*$     &     -              &     -               &     3.5     &   -   &     2     \\
363          &     27000            &    27500      &     26500      &     3.3     &    \!\!\!10 &     9     \\
384          &     28000            &     27500     &     27500      &     3.3     &    \!\!\!12  &     9     \\
469          &     29500            &     -              &     29000      &     3.5     &   0  &     1     \\
478          &     25500            &     25500     &     -               &     3.6     &   8  &     6     \\
540          &     30000            &     30000     &     30000      &     4.0     &   5  &     1     \\
553          &     24500            &     -              &     -               &     3.8     &   0  &     1     \\
572          &     25500            &     -              &     -               &     4.0     &   0  &     2     \\
616          &     29000            &     27500     &     -               &     4.4     &   0  &  \,\,\,0$^{\dagger}$ \\
623          &     28500            &     28500     &     29000      &     3.9     &   1  &     3     \\
650          &     23500$^*$     &     -              &     -               &     3.8     &   5  &     7     \\
666          &     28500            &     28000     &     -               &     4.0     &   0 &  0 \\
668          &     25000            &     25500     &     -               &     3.8     &   0  &     2     \\
673          &     25500            &     -              &     -               &     4.0     &   0  &     1     \\
692          &     29500            &     29500     &     29500       &     3.9     &   0  &     2     \\
707          &     27500            &     27000     &     -                &     4.0     &   0 &      0     \\
725          &     23500            &     24000     &     -               &     3.5      &   4  &     6     \\
740          &     22500            &     -              &     -               &     3.4      &   9  &     11     \\
748          &     29000            &     27500     &     -               &     4.1     &   4   &     5     \\
835          &     23500$^*$     &     -              &     -               &     3.8     &   0  &     2     \\
851          &     20500            &     -              &     -               &     3.4     &   -   &     0     \\
860          &     23500$^*$     &     -              &     -               &     3.8     &   0  &     6     \\
864          &     27000            &     -              &     -               &     4.1     &   0   &     1     \\
868          &     23000$^*$     &     -              &     -               &     3.6     &   6  &     6     \\
872          &     31000            &     30500     &     30500       &     4.1     &   0  &     1     \\
%879       &     18000            &     -              &     -                &     3.6   &      6     \\ Si III spectrum poor
881          &     26500            &     27000     &     -                &     3.8     &   4  &     6     \\
\hline
\end{tabular}
\end{table}
% 54 targets

\subsection{Surface Gravity} \label{s_logg}
The logarithmic surface gravity (\logg; cm s$^{-2}$) for each star was estimated by comparing theoretical and observed profiles of the hydrogen Balmer lines, H$\gamma$ and H$\delta$. H$\alpha$ and H$\beta$ were not considered as they were more affected by the nebular emission.  Automated procedures were developed to fit the theoretical spectra to the observed lines, with regions of best fit displayed by contour maps of \logg against \teff. Using the effective temperatures deduced by the methods outlined above, the gravity could be estimated. Estimates derived from the two hydrogen lines normally agreed to 0.1 dex and their averages are listed in Table \ref{t_Atm_1}. Other uncertainties may arise from, for example, errors in the normalisation of the observed spectra and uncertainty in the line broadening theory. For the former, tests using different continuum windows implied that errors in the continuum should be less than 1\%, translating into a typical error in the logarithmic gravity of less than 0.1\,dex. The uncertainty due to the latter is difficult to assess and would probably mainfest itself in a systematic error in the gravity estimates. Hence, taking these factors into consideration, a conservative uncertainty of $\pm$0.2 dex has been adopted for the gravity estimates listed in Table \ref{t_Atm_2}.

\subsection{Microturbulence} \label{s_vt}

Microturbulent velocities (\vt) have been introduced to reconcile abundance estimates from lines of different strengths for a given ionic species \citep[see, for example,][]{gra05}.  We have used the \ion{Si}{iii} triplet (4552, 4567 and 4574\,\AA)  because it is observed in almost all of our spectra and all of the lines come from the same multiplet, thereby minimising the effects of errors in the absolute oscillator strengths and non-LTE effects. This approach (method 1 in Table \ref{t_Atm_1}) has been used previously in both the Tarantula survey \citep{mce14} and a previous FLAMES-I survey \citep{duf05, hun07} and these authors noted its sensitivity to errors in the equivalent width measurements of the \ion{Si}{iii} lines, especially when the lines lie close to the linear part of the curve of growth. Indeed this was the case for four of our cooler targets where the observations (assuming typical errors of 10\% in the equivalent width estimates) did not constrain the microturbulence. The method also requires that all three lines be reliably observed, which was not always the case with the current dataset. Because of these issues, the microturbulence was also derived by requiring that the silicon abundance should be consistent with the LMC's metallicity (method 2 in Table \ref{t_Atm_1}). We adopt a value of 7.20 dex as found by \citet{hun07} using similar theoretical methods and observational data to those here; we note that this is -0.31 dex lower than the solar value given by \citet{asp09}. 

Seven targets had a maximum silicon abundance (i.e.\ for \vt = 0 \kms) below the adopted LMC value and these are identified in Table \ref{t_Atm_1}.  These targets have a mean silicon abundance of  6.96$\pm$0.08 dex. Additionally for VFTS\,119  it was not possible to remove the variation of silicon abundance estimate with line strength.  Both \citet{mce14} and \citet{hun07} found similar effects for some of their targets, with the latter discussing the possible explanations in some detail. In these instances, we have adopted the best estimate, that is zero microturbulence. For other targets an estimated uncertainty of $\pm$10\% in the equivalent widths, translated to variations of typically 3 \kms\  for both methodologies.

This uncertainty is consistent with the differences between the estimates using the two methodologies. The mean difference  (method 1 minus method 2) of the microturbulence is -0.6$\pm$2.0 \kms. Only in two cases do they differ by more than 5 \kms\ and then the estimates from the relative strength of the \ion{Si}{iii} multiplet appear inconsistent with those found for other stars with similar gravities.  However,  a conservative stochastic uncertainty of $\pm$5 \kms\ has been adopted for the values listed in Table \ref{t_Atm_2}

\begin{table}
\caption{Estimates of the atmospheric parameters and element abundances for the VFTS B-type stars.}\label{t_Atm_2}
\begin{tabular}{lrcccccr}
\hline\hline
VFTS & \vsini & \teff  &      \logg&      \vt   & \MgtoH & \NtoH  \\
     &  (\kms) &   (K)  &  (cm s$^{-2}$)   &  (\kms)  & dex & dex   \\     \hline

024     &  58~               &  27500      &     3.3     &     10     &     6.92     &     $\le$7.4     \\ %CMcE/PLD
029     &  $\leq$40~     &  27000      &     4.1     &     4       &    6.95     &     $\le$7.2     \\
050     &  $\leq$40~     &  30500      &     4.0     &     1       &     6.95     &     $\le$7.8     \\
052     &  48~               &  27500      &     3.4     &     9       &     7.03     &     6.86           \\
053     &  $\leq$40~     &  24500      &     3.5     &     4       &     7.04     &     6.71           \\
075     & 70~                &  23500      &     3.8     &     6       &     6.83     &     7.04            \\
095     &  $\leq$40~     &  31000      &     4.2     &     0       &     7.02     &     7.96           \\
111      & 80~               &  22000      &     3.3     &     0        &     6.96     &     7.44           \\
119     &  $\leq$40~     &  27500      &     4.2     &     0       &      6.88     &     $\le$7.0     \\ %RG
121     &  $\leq$40~     &  24000      &     3.6     &     0       &     6.82     &     $\le$6.9     \\
124     & $\leq$40~      &  20000      &     3.4     &     1       &     7.19     &     $\le$7.4     \\
126     & $\leq$40~      &  23500      &     3.8     &     2       &     6.76     &     $\le$6.9      \\
167     &  $\leq$40~     &  27000      &     4.0     &     0       &    6.76     &     6.96           \\ %PLD
170     & $\leq$40~      &  23000      &     3.5     &     0       &     6.87     &     $\le$6.8     \\
183     &  $\leq$40~     &  29000      &     3.5     &     4       &    6.92     &     $\le$7.6     \\

202     &  49~               &  23500      &     3.9     &     0       &     7.10     &     6.95            \\
209     &  $\leq$40~     &  27500      &     4.3     &     0       &    7.17     &     $\le$7.0     \\ %RG
214     &  $\leq$40~     &  29000      &     3.6     &     3       &    7.03     &     7.70           \\
237     & 79~                &  23000      &     3.6     &     2       &     -           &     6.92           \\
241     &   69~              &  30500      &    4.0      &     0       &     6.97      &     $\le$7.6     \\
242     &  $\leq$40~     &  29500      &     3.9     &     0       &    6.78     &     $\le$7.6     \\
273     & $\leq$40~      & 19000       &     3.8     &     5       &     6.74     &     $\le$7.7    \\
284     &  80~               &  23500      &     3.9     &     0       &     6.75     &     $\le$6.8      \\ %PLD
297     &  47~               &  23500      &     3.6     &     0       &     6.80     &     $\le$6.9     \\

308     &  74~               &  23500      &     3.8     &     0       &     7.11     &     $\le$6.7     \\
313     &  56~               &  27500      &     3.8     &     0       &     6.70     &     $\le$7.2     \\
331     &   64~              &  24000      &     3.9     &     2       &     7.03     &     $\le$7.0     \\
347     &   $\leq$40~    &  30500      &     4.0     &     0       &     7.03     &     $\le$7.4     \\
353     & 63~                &  23000      &     3.5     &     2       &     7.09     &     $\le$6.9     \\
363     &  50~               &  27000      &     3.3     &     9       &      7.08     &    7.52           \\ %CMcE/PLD
384     &  46~               &  28000      &     3.3     &     9       &     6.95     &     $\le$7.5     \\

469     &  $\leq$40~     &  29500      &     3.5     &     1       &    7.05     &     $\le$7.1     \\
478     &  64~               &  25500      &     3.6     &     6       &     6.94     &     $\le$7.2     \\

540     &  54~               &  30000      &     4.0     &     1       &     7.05     &     $\le$7.4     \\
553     &  50~               &  24500      &     3.8     &     1       &     7.02     &     6.92           \\
572     &  68~               &  25500      &     4.0     &     2       &     7.09     &     7.24           \\

616     &  $\leq$40~     &  29000      &     4.4     &     0       &    6.91     &     $\le$7.2     \\
623     &  $\leq$40~     &  28500      &     3.9     &     3       &    7.21     &     6.85           \\
650     & $\leq$40~      &  23500      &     3.8     &     7       &     6.82     &     7.02           \\
666     &  $\leq$40~     &  28500      &     4.0     &     0       &    6.94     &     $\le$7.5     \\
668     &  $\leq$40~     &  25000      &     3.8     &     2       &     7.00     &     7.13           \\
673     &  $\leq$40~     &  25500      &     4.0     &     1       &     6.98     &     $\le$7.2     \\
692     &  $\leq$40~     &  29500      &     3.9     &     2       &     7.08     &     7.69           \\

707     &  $\leq$40~     &  27500      &     4.0     &     0       &    6.96     &     $\le$6.9     \\ 
725     & $\leq$40~      &  23500      &     3.5     &     6       &     6.86     &     7.75          \\
740     & $\leq$40~      &  22500      &     3.4     &     11      &     6.92     &     7.18           \\
748     &   62 ~             &  29000      &     4.1     &     5       &     7.18     &     $\le$7.4     \\

835     &   47~              &  23500      &     3.8     &     2       &     6.86     &     6.90           \\
851     & $\leq$40~      &  20500      &     3.4     &     0       &     -           &     7.31           \\
860     &   60~              &  23500      &     3.8     &     6       &     7.13     &     6.98           \\
864     &  $\leq$40~     &  27000      &     4.1     &     1       &    7.35     &     7.46           \\ %PLD
868     & $\leq$40~      &  23000      &     3.6     &     6       &     7.12     &     7.05            \\
872     &   77~              &  31000      &     4.1     &     1       &     7.03     &     $\le$7.4     \\
881     &  $\leq$40~     &  26500      &     3.8     &     6       &     7.02     &     7.94           \\
\hline
\end{tabular}
\end{table}
% 31 <=40; 57%
% 23 40-80
% 54 targets 34 with log g>= 3.8
% 13 targets not constrained to <7.2 dex
% 7 0-40; 6 40-80
%
% 5 targets not constrained to <7.5 dex
% 4 0-40; 1 40-80

\subsection{Adopted atmospheric parameters} \label{Atm_ad}

Adopted atmospheric parameters are summarised in Table \ref{t_Atm_2}. These were taken from the silicon balance for the effective temperature apart from one target, VFTS\,241, where that estimated from  the \ion{He}{ii} spectra was used. Microturbulence estimates from the requirement of a normal silicon abundance (method 2) were adopted.

As a test of the validity of our adopted atmospheric parameters, we have also  estimated magnesium abundances (where  \MgtoH=$\log$ [Mg/H]+12) for all our sample using the \ion{Mg}{ii} doublet at 4481\,\AA. These are summarised in Tables \ref{t_Atm_2} and have  a mean of 6.98$\pm$0.14 dex. This is in good agreement agreement with the LMC baseline value of 7.05 dex found using similar methods to that adopted here \citep{hun07}; we note that our adopted LMC value is lower than the solar estimate by -0.55 dex \citep{asp09}.

\subsection{Nitrogen Abundances} \label{s_N}

Our {\sc tlusty} model atmosphere grids utilised a 51 level \ion{N}{ii} ion with 280 radiative transitions \citep{all03} and a 36 level  \ion{N}{iii} ion with 184 radiative transitions \citep{lan03}, together with the ground states of the \ion{N}{i} and \ion{N}{iv} ions. The predicted non-LTE effects for \ion{N}{ii} were relatively small compared with, for example, those for \ion{C}{ii} \citep{nie06, sig96}. For example, for a baseline nitrogen abundance, an effective temperature of 25\,000K, logarithmic gravity of 4.0 dex and a microturbulence of 5\,\kms, the equivalent width for the \ion{N}{ii} line at 3995\AA\ is 35\,m\AA\ in an LTE approximation and 43\,m\AA\ incorporating non-LTE effects. In turn, analysing the predicted non-LTE equivalent width of 43\,m\AA\ within an LTE approximation would lead to a change (increase) of the estimated nitrogen abundance of 0.15\,dex. 

Nitrogen abundances (where  \NtoH=$\log$ [N/H]+12) were estimated primarily using the singlet transition at 3995\,\AA\ as this feature was the strongest \ion{N}{ii} line in the LR02 and LR03 wavelength regions  and appeared unblended.  For stars where the singlet at 3995\,\AA\ was not visible, an upper limit on its equivalent width was estimated from the S/N of the spectroscopy. These estimates (and upper limits) are also summarised in Table \ref{t_Atm_2}.

Other \ion{N}{ii} lines were generally intrinsically weaker and additionally they were more prone to blending.  However we have searched the spectra of all of our targets for lines  that were in our {\sc tlusty} model atmosphere grids. Convincing identifications were obtained for only five targets (see Table \ref{t_Atm_2}), generally those that appear to have enhanced nitrogen abundances. These transitions are discussed below:
\\
\\
{\bf \ion{N}{ii}\,$^1$P$^{\mathrm{o}}$-$^1$D$^\mathrm{e}$\,singlet\,at\,4447\,\AA:} This feature typically has an equivalent width of approximately half that of the 3995\,\AA\ feature. Additionally it is blended with an \ion{O}{ii} line lying approximately 1.3\,\AA\ to the red. This blend could be resolved for the stars with the lowest projected velocities and fitted using two Gaussian profiles, but this inevitably affected the accuracy of the line strength estimates.
\\
\\
{\bf \ion{N}{ii}\,$^3$P$^\mathrm{o}$-$^3$P$^\mathrm{e}$\,triplet\,at\,4620\,\AA:} Transitions in this multiplet have wavelengths between 4601 and 4643\,\AA\ and  equivalent widths between approximately one third and three quarters of  that of the 3995\,\AA\ feature. This region is also rich in \ion{O}{ii} lines leading to blending problems; for example the \ion{N}{ii} line at 4643\,\AA\ is severely blended and could not be measured. Additionally the strongest transition at 4630\,\AA\ is blended with a \ion{Si}{iv} line at 0.7\,\AA\ to the red. This feature was normally weak but could become significant in our hottest targets.
\\
\\
{\bf \ion{N}{ii}\,lines\,near\,5000\,\AA:} The four features (at approximately 4994, 5001, 5005 and 5007\,\AA) arise from several triplet multiplets with the first two also being close blends. Unfortunately the last two were not useable due to very strong [\ion{O III}] nebular emission  at 5006.84\,\AA, which  also impacted on the \ion{N}{ii} blend at 5001\,\AA.  This combined with the relative weakness of the feature at 4994\,\AA \ (with an equivalent width of less than one third of that at 3995\,\AA) limited the usefulness of these lines.
\\
\\
{\bf Other\,\ion{N}{ii}\,lines:}  These were not used for a variety of reasons. The majority (at 4043, 4056, 4176, 4432, 4442, 4675, 4678, 4694, 4774, 4788, 4803\,\AA) were not strong enough for their equivalent widths to be reliably estimated even in the targets with the largest nitrogen enhancements. Others (at 4035, 4041, 4076, 4082\,\AA) were blended with \ion{O}{ii} lines. Additionally the features at 4427 and 4529\,\AA\  were blended with  \ion{Si}{iii} and  \ion{Al}{iii} lines respectively. Two \ion{N}{ii} blends from a $^3$D$^\mathrm{o}$-$^3$F$^\mathrm{e}$\ multiplet at approximately 4236 and 4241\,\AA\ were observed with equivalent widths of approximately one third of that of the 3995\,\AA\ feature. Unfortunately the upper level of this multiplet was not included in our model ion and hence these lines could not be reliably analysed.

The nitrogen abundance estimates for all the \ion{N}{ii} lines are summarised in Table \ref{t_N}, with values for the line at 3995\,\AA\ being taken directly from Table \ref{t_Atm_2}. For the two stars with the largest number of observed \ion{N}{ii} lines (VFTS 725 and 881), the different estimates are in reasonable agreement leading to sample standard deviations of less than 0.2 dex. The \ion{N}{ii} blend at 4601\,\AA\ yields a higher estimate in both stars, possibly due to blending with \ion{O}{ii} lines. For the other three targets, the sample standard deviations are smaller. Additionally, as can be seen from Table \ref{t_N}, in all cases the mean values lie within 0.1 dex of that from the line at 3995\,\AA.  In the discussion that follows, we will use the abundance estimates from this line (see Table \ref{t_Atm_2}). However, our principle conclusions would remain unchanged if we had adopted, when available, the values from Table \ref{t_N}.

\begin{table*}

\caption{Nitrogen abundance estimates from different lines for the VFTS B-type stars. The values for the line at 3995\,\AA\ have been taken directly from Table \ref{t_Atm_2}.}\label{t_N}

\begin{tabular}{lrrrrrrrrrrr}
\hline\hline
Star &  \multicolumn{8}{c}{Nitrogen Abundances (dex)} \\ 
     & 3995 & 4227 & 4447 & 4601 & 4607  & 4613  & 4621 & 4630  & 4994 & 5001 & Mean  \\   \hline  
095 & 7.96 & -       & -       & -      & -       & -       & -       & 7.87 & -       & -       & 7.92$\pm$0.06 \\
111 & 7.44 & 7.39 & -       & -      & -       & -       & -       & 7.35 & -       & -       & 7.39$\pm$0.05 \\
363 & 7.52 & -       & 7.72 & -      & -       & -       & -       & 7.62 & -       & -       & 7.62$\pm$0.10 \\
725 & 7.75 & 7.98 & 7.73 & 8.20 & 7.77 & 8.02 & 7.72 & 7.75 & 7.65 & 7.89 & 7.85$\pm$0.17 \\
881 & 7.94 & 7.90 & 7.67 & 8.07 & 7.97 & 7.93 & 7.79 & 7.83 & 7.55 & 7.96 & 7.86$\pm$0.16 \\
\hline
\end{tabular}
\end{table*}

All these abundance estimates will be affected by uncertainties in the atmospheric parameters which has been discussed by for example \citet{hun07} and \citet{fra10}. Using a similar methodology, we estimate a typical uncertainty for a nitrogen abundance estimate from uncertainties in both the atmospheric parameters and the observational data to be 0.2-0.3 dex. However due to the use of a similar methodology for all targets, we would expect the uncertainty in {\em relative} nitrogen abundances to be smaller. This would be consistent with the relatively small standard deviation for the magnesium abundance estimates found in Sect. \ref{Atm_ad}.

\section{Nitrogen abundances from the FLAMES-I survey}

As discussed in Sect.\ \ref{s_obs}, we have re-evaluated the nitrogen abundance estimates for the apparently single LMC targets \citep{eva06} in the FLAMES-I survey found by  \citet{hun07, hun08b, hun09a} and \citet{tru07}. Our approach was to adopt their atmospheric parameters but to use our re-reduced spectroscopy of the \ion{N}{ii} line at 3995\,\AA. Our target selection followed the same criteria as adopted here, viz. B-type targets (excluding supergiants) with no evidence of binarity. This led to the identification of 34 targets\footnote{Further consideration of the FLAMES spectroscopy led to two changes compared with the binarity classification in \citet{eva06}. NGC2004/91 was reclassified as apparently single and NGC2004/97 as a binary candidate. Additionally N11/101 was not analysed due to the relatively poor quality of its spectroscopy.}, including two for which it had not been possible to obtain nitrogen abundance estimates in the original survey. The projected rotational velocity, atmospheric parameters and revised nitrogen abundance estimates from the \ion{N}{ii} at 3995\,\AA\ are summarised in Table \ref{t_FLAMESI}.

\begin{table}
\caption{Estimates of the atmospheric parameters, projected rotational velocities and nitrogen abundances for narrow lined B-type targets from the FLAMES-I survey for the LMC clusters NGC\,2004 (designated 2004) and N11.}\label{t_FLAMESI}
\begin{tabular}{lllccccr}
\hline\hline
Star  & \teff  &      \logg&      \vt  & \vsini & \NtoH\ \\
     &      (K)   &   (cm s$^{-2}$)   &  (\kms) & (\kms)  & dex \\     \hline
2004/36        &   22870     &   3.35    &   7    &  42   &  7.27 \\
2004/42        &   20980     &   3.45    &   2    &  42   &  6.87 \\
2004/43        &   22950     &   3.50    &   7    &  24   &  7.13 \\
2004/46        &   26090     &   3.85    &   2    &  32    &  7.55 \\
2004/51        &   21700     &   3.40    &   5    &  70   &  6.87 \\
2004/53        &   31500     &  4.15     &   6    &  7     &  7.64 \\
2004/61        &   20990     &   3.35    &   1    &  40   &  6.85 \\
2004/64        &   25900     &   3.70    &   6    &  28   &  7.59 \\
2004/68        &   20450     &   3.65    &   1    &  62   &  6.95 \\
2004/70        &   27400     &   3.90    &   4    &  46   &  7.56 \\
2004/73        &   23000     &   3.65    &   1    &  37   &  6.92 \\
2004/76        &   20450     &   3.70    &   1    &  37   &  6.88 \\
2004/84        &   27395     &   4.00    &   3    &  36   &  7.15 \\
2004/86        &   21700     &  3.85     &   6    &  14   &  7.07 \\
2004/87        &   25700     &  4.40     &   0    &  35   &  6.90 \\
2004/91        &   26520     &  4.10     &   0    &  40   &  7.12  \\ % Redesignated as single
%NGC2004/97   &21700      &   3.60	     &2       &  62    &7.04 \\ Redesignated as a binary
2004/103      &  21500      &  3.85     &  1     &  35    &  6.84 \\
2004/106      &  21700      &  3.50     &  2     &  41    &  6.87 \\
2004/111       &  20450      &  3.30     &  2     &  55    &  6.94 \\
2004/112       &  21700      &  3.70     &  1     &  72    &  6.81 \\
2004/114       &  21700      &  3.60     &  2     &  59    &  6.92 \\
2004/116       &  21700      &  3.55     &  3     &  63    &  6.93 \\
2004/117       &  21700       &  3.60     &  4     & 75     & 7.07 \\
2004/119       &  23210       &  3.75     &  0     & 15     & 6.86 \\

N11/69           & 24300         &  3.30    &  11    & 80     &  6.83 \\
N11/70           & 19500         &  3.30    &   0     & 62     &  6.79 \\
N11/72           & 28800         &  3.75    &   5     &15      &  7.43 \\
N11/79           & 32500         &  4.30    &   0     &38      &  7.02 \\
N11/86           & 26800         &  4.25    &   0     & 75     &  6.97 \\
N11/93           & 19500          &  3.30   &    1    & 73     &  6.78 \\
N11/100         & 29700         &   4.15   &   4     & 30     &  7.78 \\
%N11/101      & 29800         &   3.95   &   8     & 70     &  $\leq$7.0 \\ Excluded as small ew
N11/106         & 31200         &   4.00   &   5	 & 25     &  6.94 \\
N11/110         & 23100         &   3.25   &   7    & 25     &  7.43  \\
N11/115         & 24150         &   3.65   &   1	 & 53     &  6.91 \\		
\hline
\end{tabular}
\end{table}
% 17 <=40 50%
% 16 40-80
% 34 targets

\begin{figure}
\epsfig{file=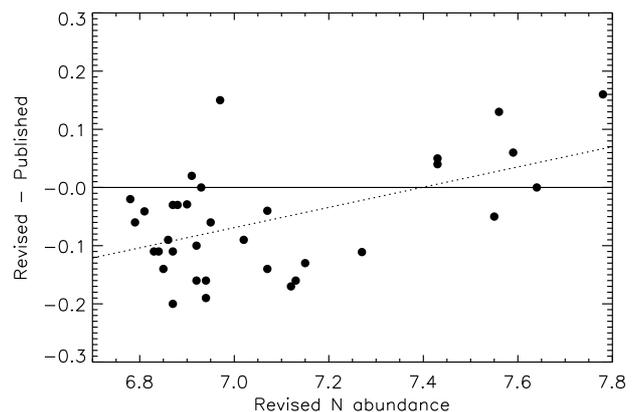,width=9cm, angle=0} 

\caption{Comparison of revised nitrogen abundance estimates for the FLAMES-I survey listed in Table \ref{t_FLAMESI} with those found by  \citet{hun07, hun08b, hun09a} and \citet{tru07}. The abscissa is our revised nitrogen abundance, \NtoH, whilst the ordinate is the difference between the revised value and that found previously. A least squares linear fit to the data is shown as a dotted line.}
\label{N_ab}
\end{figure} 

Fig. \ref{N_ab} illustrates the differences between our revised nitrogen abundance estimates and those found previously. Although the two sets ore in reasonable overall agreement (with a mean difference of -0.06$\pm$0.11 dex), there appears to be a trend  with nitrogen abundance estimates. An unweighted linear least squares fit also (shown in Fig. \ref{N_ab}) has a slope of 0.083$\pm$0.066 and is therefore not significant at a 95\% confidence level. 

Unfortunately without access to the original equivalent width data, it is difficult to identify the sources of any differences. Generally the smaller original abundance estimates were based on one or two lines and hence the differences may reflect different equivalent width estimates for the \ion{N}{ii} line at 3995\AA. As discussed in Sect. \ref{s_obs}, we have re-reduced the FLAMES-I spectroscopy and would expect that the current equivalent width estimates will be the more reliable. The larger original nitrogen abundance estimates were based on typically 5-6 lines and any differences may in part reflect this. In Sect. \ref{d_obs}, we consider what changes the use of the original nitrogen abundance would have made to our discussion.

%----------------------------------------------------------------------
\section{Discussion}\label{s_discussion}

\subsection{Stellar parameters} \label{s_Atm}

\begin{figure}
\epsfig{file=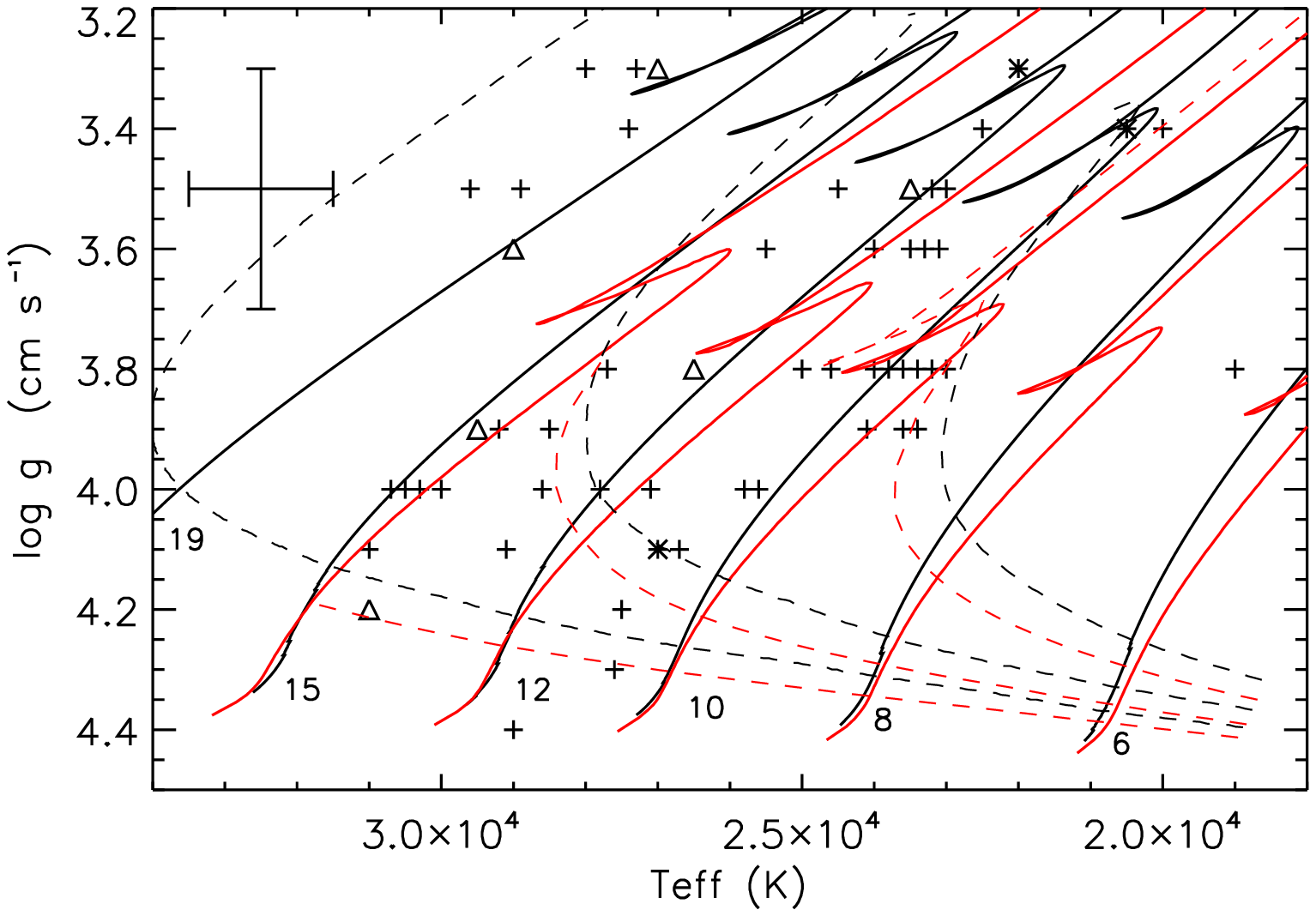,width=9cm, angle=0} 
\epsfig{file=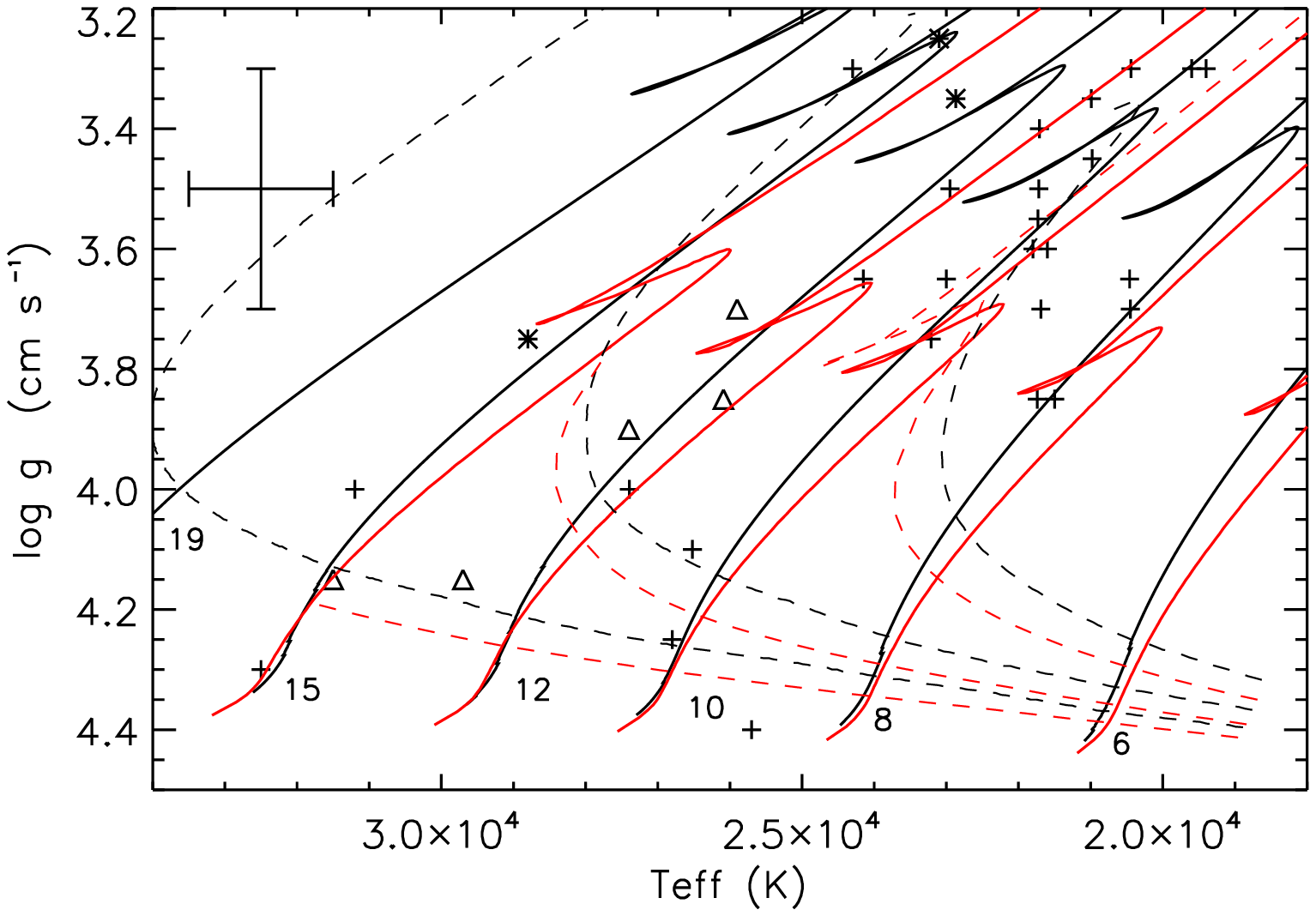,width=9cm, angle=0} 

\caption{Estimates of the atmospheric parameters for the targets from the VFTS (upper) and FLAMES-I  survey  (lower); some targets have been moved slightly in effective temperature to improve clarity. Targets with an estimated nitrogen abundance, 7.2$\le$\NtoH$<$7.5 dex, are  shown as asterisks and those with \NtoH$\ge$7.5 dex  are shown as triangles; all other targets are shown as crosses. Evolutionary models of \citet{bro11a} (solid black lines) and \citet{geo13a} (solid red lines) are shown for zero initial rotational velocity together with the initial mass (in units of the solar mass). Isochrones (dashed lines) are shown as dotted lines with the same colour selection for ages of 5, 10 and 20~Myr. Typical uncertainties in the effective temperatures and gravity estimates are also shown.}

\label{atm_par}
\end{figure}

The estimates of the effective temperatures and surface gravities for the VFTS and FLAMES-I samples are illustrated in Fig. \ref{atm_par}. They cover similar ranges in atmospheric parameters, viz. 18\,000\,$\la$\,\teff\,$\la$\,32\,000 K and 3.2\,$\la$\,\logg\,$\la$\,4.4 dex, with the lower limit in gravity following from the criteria for target selection. Neither sample contains targets with low effective temperatures and large gravities, as these would be relatively faint and would lie below the apparent magnitude limit for observation. The estimated microturbulences of the two samples are also similar, covering the same range  of 0--11 \kms\ and having median and mean values of 1.5 and 2.6 \kms\ (VFTS) and 2.5 and 3.3 \kms\ (FLAMES-I).

The estimated projected rotational velocities of our two samples lie between 0--80 \kms\ due to the selection criteria and in the subsequent discussion two ranges are considered, viz, \vsini\,$\leq$40 \kms\ and  40\,$<$\,\vsini\,$\leq$\,80 \kms. The VFTS estimates of \citet{duf12} adopted a Fourier Transform methodology \citep[see,][for details]{car33, sim07}, whilst the FLAMES-I measurements of \citet{hun08a} were based on profile fitting  of rotationally broadened theoretical spectra to the observations. 

As discussed by \citet{sim14}, both methods can be  subject to errors due  the effects of other broadening mechanisms, including macroturbulence and microturbulence. They concluded that for {\em O-type dwarfs and B-type supergiants}, estimates for targets with \vsini\,$\le$\,120\,\kms\ could be overestimated by $\sim$25$\pm20$\,\kms. Recently \citet{sim17} have discussed in detail the non-rotational broadening component in OB-type stars. They again find that their {\em O-type stellar and B-type supergiant samples} are 'dominated by stars with a remarkable non-rotational broadening component'; by contrast, in the B-type main sequence domain, the macroturbulence estimates are generally smaller, although the magnitude of any uncertainty in the projected rotational velocity is not quantified. However inspection of Fig.5 of \citet{sim17}, implies that the  macroturbulence will be typically 20-30 \kms\ in our targets compared with values of 100 \kms\ or more that can be found in O-type dwarfs and OB-type supergiants.

The effects of these uncertainties on our samples are expected to be limited for three reasons. Firstly the results of \citet{sim17} imply that the effects of non-rotational broadening components are likely to be smaller in our sample than in the O-type stellar and B-type supergiants discussed by \citet{sim14}. Secondly, we have not used specific \vsini\ estimates but rather placed targets into \vsini\  bins. Thirdly for our sample and especially the cohort with \vsini\,$\leq$\,40\,\kms, the intrinsic narrowness of the metal line spectra requires a small degree of rotational broadening. In Sect.\ \ref{d_obs}, we discuss the implications of possible uncertainties in our \vsini\  estimates.

\subsection{Nitrogen abundances} \label{s_GC}

 As discussed by \citet{gar17}, the LMC baseline nitrogen abundance has been estimated from both \ion{H}{ii} regions and early-type stars. For the former,  \citet{kur98}  and  \citet{gar99} found 6.92 and 6.90 dex respectively. \citet{kor02, kor05}, \citet{hun07} and \citet{tru07} found a range of nitrogen abundances from their analyses of B-type stellar spectra, which they attributed to different degrees of enrichment by rotational mixing. Their lowest estimates implied baseline nitrogen abundances of  6.95 dex (1 star), 6.90 dex (5 stars) and  6.88 dex (4 stars) respectively. These different studies are in good agreement and a value of 6.9~dex has been adopted here, which is -0.93 dex lower than the solar estimate \citep{asp09}

The majority of targets appear to have nitrogen abundances close to this baseline, with approximately 60\% of the 54 VFTS targets  and 75\%  of the 34 FLAMES-I targets having enhancements of less than 0.3 dex. Additionally for the VFTS sample, an additional 20\% of the sample have upper limits greater than 7.2 dex but could again either have modest or no nitrogen enhancements. Hence  approximately three quarters of each sample may have nitrogen enhancements of less than a factor of two. This would, in turn be consistent with them having relatively small rotational velocities and having evolved without any significant rotational mixing  \citep[see, for example,][]{mae87, heg00b, mae09,  fri10}.

The two targets, VFTS\,167 and 469 whose spectroscopy was probably affected by fibre cross-contaminations (see Sect.\ \ref{s_obs}), have atmospheric parameters that are consistent with both their spectral types and other VFTS targets. Additionally they appear to have near baseline nitrogen abundances. Although these results should be treated with some caution, it appears unlikely that they belong to the subset of targets with large nitrogen enhancements discussed below.

Two targets, VFTS\,237 and 835, have been identified as Be-type stars and might be expected to show the effects of rotational mixing due to rotational velocities near to critical velocities having been inferred \citep{tow04, cra05, ahm17} for such objects. However both have near baseline nitrogen abundance estimates (6.92 and 6.90 dex respectively) implying that little rotational mixing has occurred. Our analysis did not make any allowance for the effects of light from a circumstellar disc. \citet{dun11} analysed spectroscopy of 30 Be-type stars in the Magellanic Clouds and estimated a wide range of disc contributions, $\sim$0-50\%. In turn this led to increases in their nitrogen abundances in the range $\sim$0-0.6 dex with a median increase of 0.2 dex. Hence a nitrogen enhancement in these two objects cannot be discounted. However  near baseline nitrogen abundances have previously been found for Be-type stars in both the Galaxy \citep{ahm17} and the Magellanic Clouds \citep{len05, dun11}, with \citet{ahm17} having recently discussed possible explanations.

The remaining targets show significant enhancements in nitrogen of up to 1.0 dex. For both samples,  $\sim$20-25\% show an enhancement of greater than 0.3 dex with $\sim$10-15\% having enhancements of greater than 0.6 dex. These could be rapidly rotating stars that have undergone rotational mixing and are being observed at small angles of inclination. However such cases would be expected to be relatively rare. For example, assuming random axes of inclination, a star with an equatorial velocity of 300 \kms\ would have a value of \vsini\,$\leq$40 \kms\ only about 1\% of the time. Additionally the probability that such  a star would be observed with 40\,$<$\,\vsini\,$\leq$\,80 \kms\ is approximately four times that of it being observed with \vsini\,$\leq$\,40 \kms. Similar ratios would be found for other values of the rotational velocity and just reflects the greater solid angle of inclination available to stars in the larger projected rotational velocity bin. By contrast, the nitrogen rich targets in our samples are more likely to have \vsini\,$\leq$\,40 \kms\ than 40\,$<$\,\vsini\,$\leq$\,80 \kms, as can be seen from Table  \ref{t_N_sim}.  In Sect. \ref{d_sim}, we return to these objects and undertake simulations to investigate whether all the nitrogen enhanced stars in our sample could indeed be fast rotators observed at small angles of inclination.

\begin{figure*}
\psfig{file=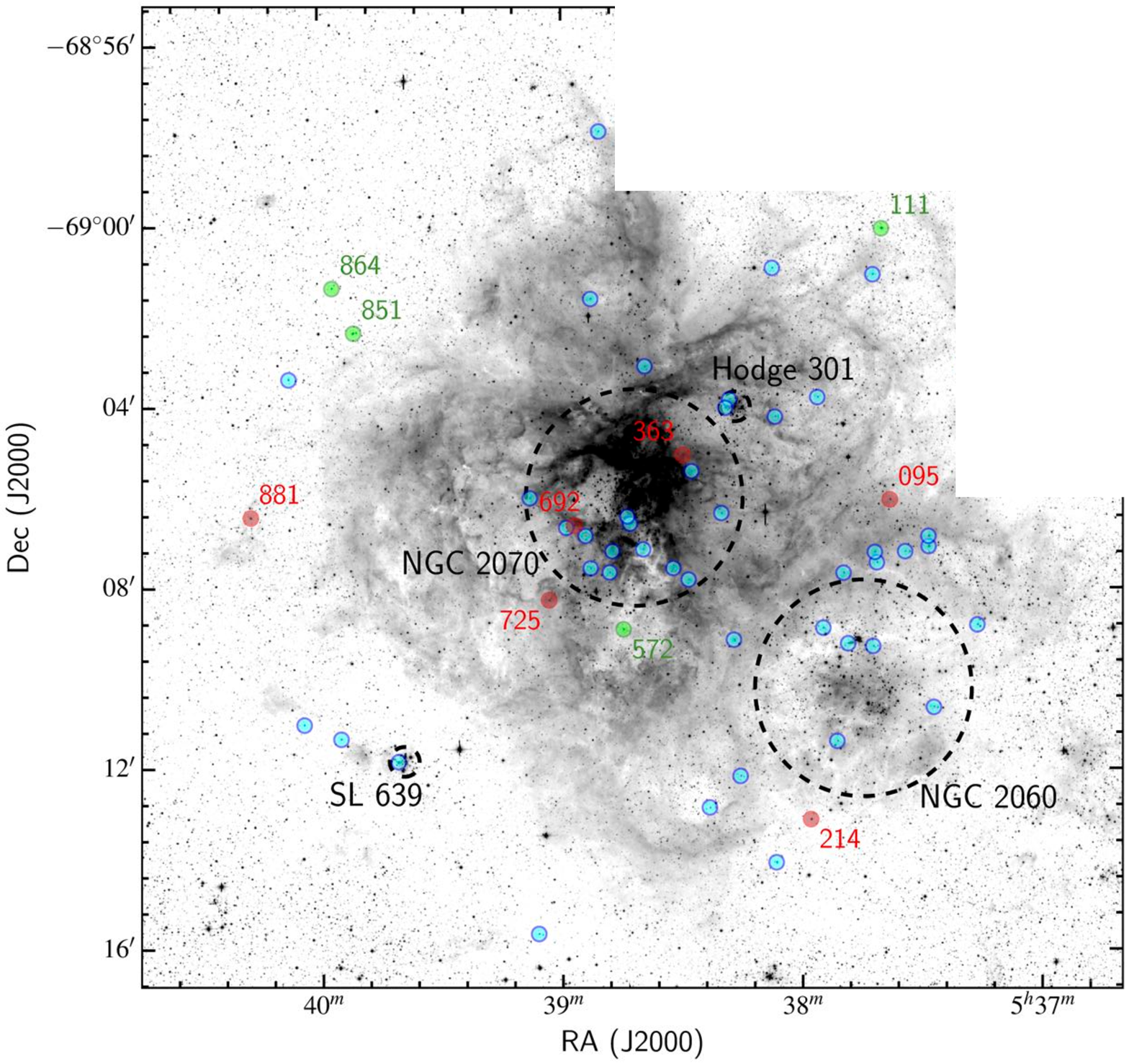,width=19
cm, angle=0} 

\caption{Spatial distribution of the VFTS targets in 30 Doradus.  Targets with 7.2$\le$\NtoH$<$7.5 dex, are  shown in green and those with \NtoH$\ge$7.5 dex  in red; other targets are shown in blue. The spatial extents of NGC 2070, NGC 2060, SL 639, and Hodge 301 \citep{eva15} are indicated by the overlaid dashed circles. The underlying image is from a V-band mosaic taken with the ESO Wide Field Imager on the 2.2 m telescope at La Silla.}
\label{spatial}
\end{figure*}

We have investigated the other observed characteristics of our nitrogen enhanced stars to assess whether they differ from those of the rest of the sample. Fig.\ \ref{spatial} shows the spatial distribution of all of our VFTS targets, together with the position and extent of the larger clusters NGC 2070 and NGC 2060 and the smaller clusters, Hodge 301 and SL639; the sizes of the clusters should be treated as representative and follow the definitions of \citet[][Table 4]{eva15}. The nitrogen enriched targets are distributed across the 30 Doradus region with no evidence for clustering within any cluster. Adopting the cluster extents of \citet{eva15}, there are 21 cluster and 33 field VFTS stars. All four targets with 7.2$\le$\NtoH$<$7.5 dex lie in the field whilst those with  \NtoH$\ge$7.5 dex consist of four field and two cluster stars. Hence of the ten nitrogen enriched stars, only two lie in a cluster implying a possible excess of field stars. However simple binomial statistics, imply that this excess is not significant at the 10\% level for either the total sample or those in the two abundance ranges.  

The dynamics of our sample can also be investigated using the radial velocity measurements summarised in Table \ref{t_Targets}. The mean radial velocity for the whole sample is 270.3$\pm$16.6\,\kms\ (54 targets), whilst the 10 nitrogen enriched stars and the remaining 44 stars have values of $270.2\pm 18.9$\kms\ and $270.3\pm 16.3$\kms\ respectively. The latter two means are very similar whilst an F-test of the corresponding variances showed that they were not significantly different at even the 20\% level. Hence we conclude that the radial velocity distributions of the two subgroups appear similar within the limitation of the sample sizes.

A comparison of the atmospheric parameters also reveals no significant differences. For example, the whole VFTS sample have a median and mean effective temperature of 27\,000 and 25\,865\,K, while the values for the targets with \NtoH$\ge$7.2 dex are 26\,500 and 26056\,K. The corresponding statistics for the gravity are 3.80/3.77 dex (whole sample)  and 3.75/3.74 dex (nitrogen enriched) and for the microturbulence 1.5/2.6 \kms\ (whole sample)  and 2.0/2.2 \kms\ (nitrogen enriched).

A helium to hydrogen abundance ratio by number of 0.1 was assumed in our calculations of theoretical spectra (see Sect.\ \ref{s_meth}). The LMC evolutionary models of \citet{bro11a} appropriate to B-type stars indicate that even for a nitrogen enhancement of 1.0 dex (which is the largest enhancement observed in our sample -- see Tables \ref{t_Atm_2} and \ref{t_FLAMESI}), the change in the helium abundance is typically only  0.03 dex. Hence our assumption of a normal helium abundance is unlikely to be a significant source of error. We have checked this by comparing theoretical and observed \ion{He}{i} profiles for targets with the largest nitrogen enhancements. This comparison was impacted by significant nebula emission especially for lines in the (3d)$^3$D-(4f)$^3$F series but generally yielded good agreement between observation and theory for our adopted helium abundance. 

\begin{figure}
\epsfig{file=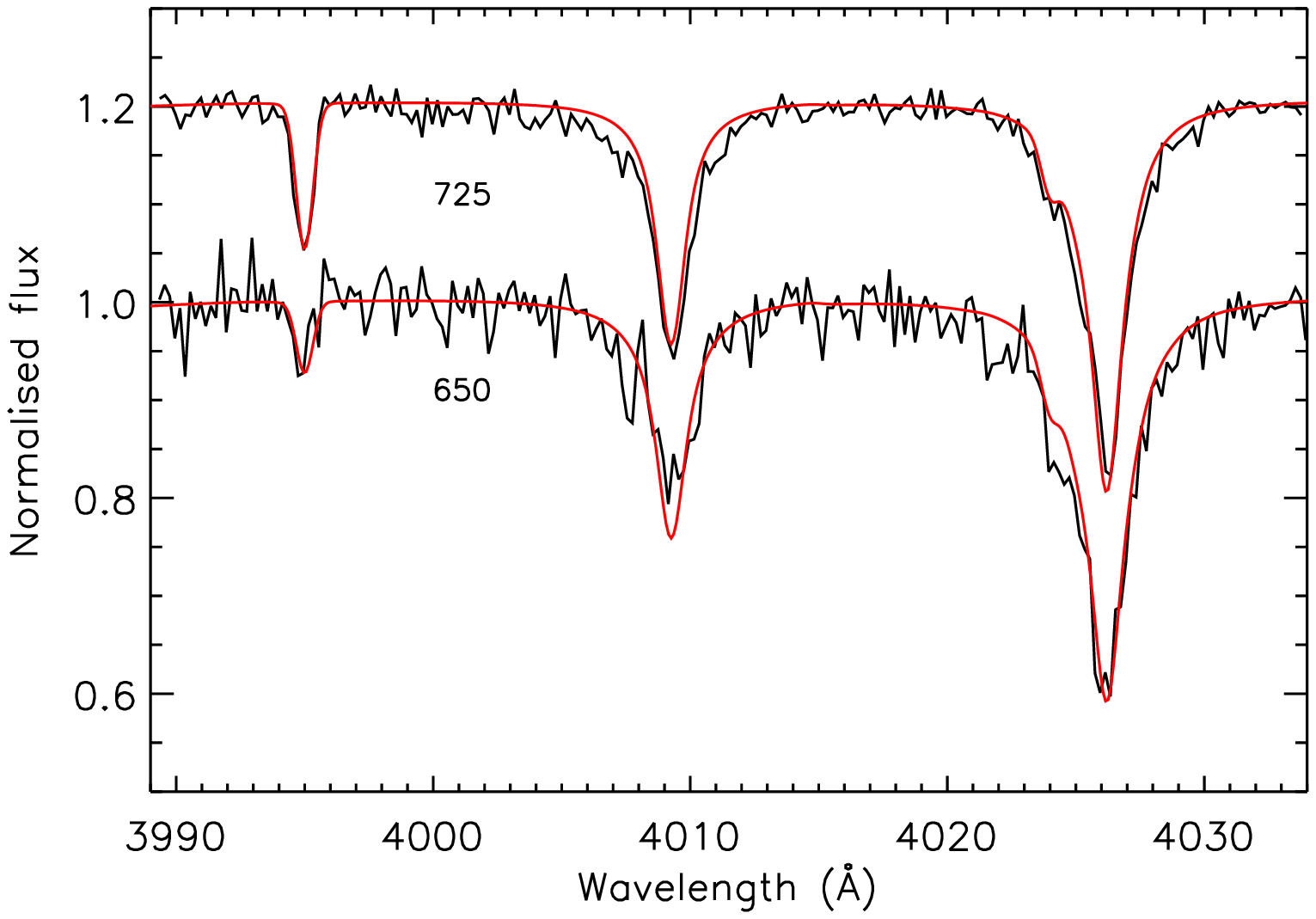,width=9cm, angle=0} 

\caption{Observed and theoretical spectra for the wavelength region near 4000\,\AA\ in two stars, VFTS\,650 and 725. Atmospheric parameters and nitrogen abundances were taken from Table \ref{t_Atm_2} for the theoretical spectra. The normalised spectra for VFTS\,725 have been shifted upwards to improve clarity}
\label{hei}
\end{figure}

This is illustrated by observed and theoretical spectra of the wavelength region 3980-4035\,\AA\ shown in Fig. \ref{hei} for VFTS\,650 and 725, which have similar atmospheric parameters. The observed spectra have been corrected using the radial velocity estimates listed in Table \ref{t_Targets}, while the theoretical spectra have been interpolated from our grid to be consistent with the atmospheric parameters and nitrogen abundances listed in Table \ref{t_Atm_2}. The latter have also been convolved with a Gaussian function to allow for instrumental broadening; as both stars have estimated projected rotational velocities, $\le$\,40 \kms, no correction was made for rotational broadening. For the two diffuse \ion{He}{i} lines at approximately 4009 and 4026\,\AA, the agreement is excellent although VFTS\,650 has a near baseline nitrogen abundance whilst VFTS\,725 has the second highest nitrogen abundance in the VFTS sample. The agreement between theory and observation for the \ion{N}{ii} at 3995\,\AA\ is also good and supports our use of equivalent widths to estimate element abundances. The observed spectra also illustrate the range of observed S/N with VFTS\,650 (S/N$\sim$60) and 725 (S/N$\sim$120) having estimates towards to the lower and upper ends of that observed.

For the nitrogen enriched stars, changes in the carbon and oxygen abundances would also be expected. \citet{mae14} argued that comparison of the relative N/C and N/O abundances would provide a powerful diagnostic of rotational mixing. The LMC models of \citet{bro11a} appropriate to B-type stars indicate that for a nitrogen enhancement of $\sim$1.0 dex, the carbon and oxygen abundances would be decreased by -0.22--0.35 dex and -0.08--0.10 dex respectively. 

Unfortunately the strongest \ion{C}{ii}\ line in the blue spectra of B-type stars is the doublet at 4267\,\AA, which is badly affected by nebular emission. Due to the sky and stellar fibres being spatially separated, it was not possible to reliably remove this emission from the VFTS spectroscopy. Indeed as discussed by \citet{mce14} this can lead to the reduced spectra having spurious narrow emission or absorption features. This is especially serious for targets with small projected rotational velocities, where the width of the nebular emission is comparable with that of the stellar absorption lines. No other \ion{C}{ii} lines were reliably observed in the VFTS spectra and hence reliable carbon abundances could not be deduced. The FLAMES-I spectroscopy may also be affected by nebular emission although this might be expected to be smaller especially in the older cluster, NGC2004. However without a reliable method to remove nebular emission, the approach advocated by \cite{mae14} would appear to be of limited utility for multi-object fibre spectroscopy of low \vsini\ early-type stars.

Both the VFTS and FLAMES-I data have rich \ion{O}{ii} spectra. However the predicted changes in the oxygen abundance (of less than 0.1 dex) are too small to provide a useful constraint, given the quality of the observational data. This is confirmed by the original FLAMES-I analyses \citep{hun07, hun08b, hun09a, tru07}, where for the targets considered here an oxygen abundance of 8.38$\pm$0.15 dex was found \citep[which is lower than the solar estimate by -0.31 dex,][]{asp09}. The standard error is similar to that found in Sect. \ref{Atm_ad} for the magnesium abundance where no variations within our sample are expected. Additionally for the five FLAMES-I stars with a nitrogen enhancement of more than 0.6 dex, the mean oxygen abundance was 8.40$\pm$0.08 consistent with that found for all the targets.

\begin{table}

\caption{Estimates of the initial rotational velocity, \vi,  required to achieve nitrogen enhancements of greater than 0.3 dex (\NtoH$\geq$7.2 dex) and 0.6 dex (\NtoH$\geq$7.5 dex) whilst maintaining a gravity, \logg$\geq$ 3.3 dex. Also listed are the minimum rotational velocities, \ve, predicted by the models during this evolutionary phase. Initial masses (\Mi) are in units of the solar mass, \Msun. The values characterised as 'Adopted' were used in the simulations discussed in Sects. \ref{d_sim}-\ref{d_freq}.}\label{t_v_e}

\begin{center}
\begin{tabular}{lcccc}
\hline\hline
\Mi/\Msun &  \multicolumn{2}{c}{\NtoH$\geq$7.2 dex} &  \multicolumn{2}{c}{\NtoH$\geq$7.5 dex} \\
                   & \vi & \ve & \vi & \ve \\
                   &\kms&\kms&\kms&\kms \\   \hline  
\,\,\,5 & 180 & 142 & 258 & 219\\
\,\,\,8 & 161 & 132 & 230 & 207\\
12     &  163 & 140 & 219 & 200\\
16     & 144 & 130 & 217 & 203 \\ 
Adopted & 145 & 135 & 220 & 205 \\
\hline
\end{tabular}
\end{center}
\end{table}

\begin{table*}

\caption{Observed and predicted numbers of nitrogen rich targets with enhancements of greater than 0.3 dex (\NtoH$\geq$7.2 dex) and 0.6 dex (\NtoH$\geq$7.5 dex). The Monte-Carlo simulations are designated as 'MC'. When two number of targets, $N$, are listed, they refer to simulations tailored to the different nitrogen abundance limits as discussed in the text. The rotational velocity distribution, $f(v_\mathrm{e})$, is either from \citet[][VFTS]{duf12} or from the deconvolution illustrated in Fig. \ref{v_e} (FLAMES-I). Simulations are provided for two ranges of projected rotational velocity, \vsini, viz. 0-40 \kms\ and 40-80 \kms.}\label{t_N_sim}

\begin{center}
\begin{tabular}{lrr|rr|rrrrrrr}
\hline\hline
Sample  &  $N$ & $f(v_e)$  &\multicolumn{2}{c|}{\NtoH$\geq$7.2} &  \multicolumn{2}{c}{\NtoH$\geq$7.5} \\
 \vsini\ (\kms)   &&& 0-40 & 40-80 & 0-40 & 40-80 \\   \hline  
&&&&&\\
\multicolumn{2}{l}{\underline{VFTS sample}} &&&& \\
&&&&&\\ 
Observed & 255 & -  & 7      &  3      & 5      & 1 \\
&&&&&\\
%Int       & 255 & VFTS & 2.28    &  7.10   & 1.45   & 4.45\\
MC      & 255 & VFTS & 2.30$\pm$1.50    & 7.17$\pm$2.63 & 1.45$\pm$1.18 & 4.46$\pm$2.08\\ 
MC Prob (\%)   & 255 & VFTS & 0.8 & 7 & 2 & 6 \\ 
%MC Prob (\%)   & 255 & VFTS & 0.80 & 7.01 & 1.56 & 6.09 \\ 
&&&&&\\                               
%Int       & 220/188 & VFTS & 1.97    &  5.29   & 1.25   & 3.32\\
MC      & 220/188 & VFTS & 1.98$\pm$1.14    & 5.29$\pm$2.27 & 1.25$\pm$1.11 & 3.32$\pm$1.80\\ 
MC Prob (\%)   & 220/188 & VFTS & 0.4 & 22 & 0.9 & 16 \\                                
%MC Prob (\%)   & 220/188 & VFTS & 0.43 & 22.43 & 0.89 & 15.82 \\                                
&&&&&\\ 
&&&&&\\ 
\multicolumn{2}{l}{\underline{FLAMES-I sample}}&&&&& \\
&&&&&\\ Observed  &103 & - & 6 & 2 & 4 & 1\\
&&&&&\\
%Int          & 103 & VFTS & 0.92 & 2.87 & 0.59 & 1.80 \\
MC             & 103 & VFTS & 0.92$\pm$0.96    & 2.89$\pm$1.69 & 0.58$\pm$0.76 & 1.82$\pm$1.33\\
MC Prob(\%)     & 103 & VFTS & 0.03 & 45  & 0.3 & 46\\
%MC Prob(\%)     & 103 & VFTS & 0.03 & 44.74  & 0.28 & 46.15\\
&&&&&\\
%Int           & 73 & FLAMES-I & 0.58 & 1.82 & 0.26 & 0.81 \\
MC             & 73 & FLAMES-I & 0.57$\pm$0.76    & 1.82$\pm$1.33 & 0.26$\pm$0.51 & 0.79$\pm$0.89\\
MC Prob(\%)     & 73 & FLAMES-I & 0.01 & 55  & 0.04 & 56\\
 \hline
\end{tabular}
\end{center}
\end{table*}

\subsection{Estimation of numbers of nitrogen enriched targets} \label{d_sim}

As discussed above, it is unclear whether the targets with the largest nitrogen enhancements in our samples could have evolved as single stars undergoing rotational mixing. We have therefore undertaken simulations to estimate the number of rotationally mixed stars that we might expect to find in our sample.  We have considered two thresholds for nitrogen enrichment, viz. 0.3 dex (corresponding to a nitrogen abundance, \NtoH$>$7.2 dex) and 0.6 dex (\NtoH$>$7.5 dex). The former was chosen so that any enhancement would be larger than the estimated errors in the nitrogen abundances, whilst the larger was chosen in an attempt to isolate the highly enriched targets, at least in terms of those lying near to the main sequence. As discussed by \citet{mce14}, B-type supergiants (both apparently single stars and those in binary systems) in 30 Doradus can  exhibit larger nitrogen abundances, \NtoH, of up to 8.3 dex corresponding to an enhancement of 1.4 dex; however these stars have probably evolved from the O-type main sequence. The number of stars fulfilling these criteria in both the VFTS and FLAMES-I samples are summarised in Table \ref{t_N_sim}.

We have used the LMC grid of models of \citet{bro11a} to estimate the initial stellar rotational velocities, \vi, that would be required to obtain such enrichments. The core hydrogen burning phase, where the surface gravity, \logg $\geq$ 3.3 dex, was considered in order to be consistent with the gravities estimated for our samples (see Tables \ref{t_Atm_2} and \ref{t_FLAMESI}). We have also found the {\em minimum} rotational velocity, \ve, that a star would have during this evolutionary phase; this velocity was normally found at a gravity, \logg$\simeq$3.3 dex. The initial masses were limited to 5-16\Msun\ as the ZAMS effective temperatures in this evolutionary phase range from approximately 20\,000 to 33\,000K, compatible with our estimated effective temperatures, again listed in Tables \ref{t_Atm_2} and \ref{t_FLAMESI}.

For each initial mass, we have quadratically interpolated between models with different initial (\vi) rotational velocities to estimate both the initial (\vi) and the minimum rotational velocity (\ve\ - hereafter designated as `the current rotational velocity') that would achieve a nitrogen enhancement of at least either 0.3 or 0.6 dex whilst the surface gravity, \logg$\geq$ 3.3 dex. These estimates are summarised in Table \ref{t_v_e}. Given that the grid spacing in the initial rotational velocity was  typically 60 \kms\ (or less), we would expect that the interpolation would introduce errors of less than 20 \kms. The ranges of the estimated initial and current velocities for different masses are approximately 20\% and 15\% respectively. For our simulations, we have taken a conservative approach and adopted velocities (see Table \ref{t_v_e}) at the lower ends of our estimates.  In turn this will probably lead to an overestimate in the number of targets with enhanced nitrogen abundances and small projected rotational velocities in our simulations. This overestimation will be re-enforced by the adoption of a lower  gravity limit, \logg$\geq$3.3 dex whilst our estimated gravities are generally larger; for example over half the VFTS sample have estimated gravities, \logg $\geq 3.8$\ dex.

As discussed by \citet{gra05}, for a random distribution of rotation axes, the probability distribution for observing an angle of inclination, $i$, is:
\begin{equation}\label{e_1}
P(i)\ di = \sin i\ di
\end{equation}

\noindent
the probability of observing an angle of inclination, $i\leq i_0$\ is then given by:
\begin{equation}\label{e_2}
P(i\leq i_0) =1-\cos i_0
\end{equation}

\noindent
For a sample of $N$\ targets with a normalised rotational velocity distribution, $f(v_e)$, the number that will have a projected rotational velocity less than $v'$  is given by:
\begin{equation}\label{e_3}
N(v_e\sin i \leq v')=N \left [\int^{v'}_{0}f(v_e) dv_e + \int^{\infty}_{v'} f(v_e) P\left (i \leq i_0\right ) dv_e \right ]
\end{equation}

\noindent or
\begin{equation}\label{e_4}
N(v_e\sin i \leq v')=N \left [\int^{v'}_{0}f(v_e) dv_e + \int^{\infty}_{v'} f(v_e) \left (1- \cos i_0 \right )dv_e   \right ]
\end{equation}

\noindent where
\begin{equation}\label{e_5}
i_0= \sin^{-1}\left(\frac{v'}{v_e}\right )
\end{equation}

\noindent The number of targets that have both  \vsini\,$\leq v'$ and a nitrogen enhancement, \NtoH$\geq$\NtoHL, can then be estimated by changing the lower limits of integration in equation \ref{e_4} to the rotational velocities, \ve, summarised in Table \ref{t_v_e}. This will normally eliminate the first term on the right hand side of equation \ref{e_4} leading to:

\noindent 
\begin{equation}\label{e_6}
\begin{split}
N(v_e\sin i \leq v'; \epsilon \geq \epsilon_\mathrm{0})=N\int^{\infty}_{v_\mathrm{e}} f(v_e)\left [1- \cos i_0 \right ]dv_e
\end{split}
\end{equation}

\begin{figure}
\epsfig{file=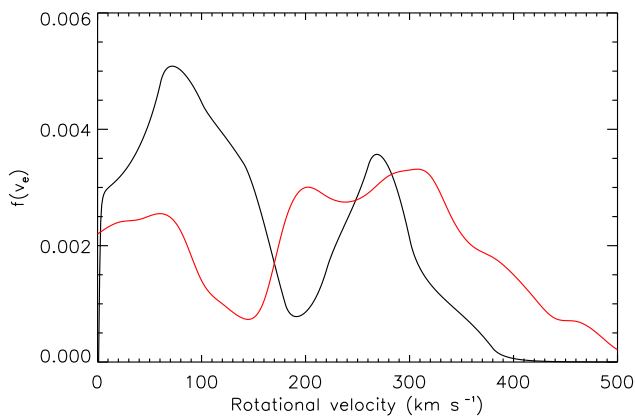,width=9cm, angle=0} 

\caption{The de-convolved rotational velocity distribution, $f(v_e)$, for the LMC FLAMES-I sample of single B-type targets with luminosity classes V to III (black line) compared with that for the equivalent VFTS sample taken from \citet[][red line]{duf12}.}
\label{v_e}
\end{figure}

The rotational velocity distribution,  $f(v_e)$, for the VFTS B-type single star sample has been deduced by \citet[][see their Table 6]{duf12}. \citet{hun08a} assumed a single Gaussian distribution for the rotational velocities in the FLAMES-I single B-type star sample and fitted this to the observed \vsini\ distribution. We have taken their observed values and used the deconvolution methodology of \citet{luc74} as described in \citet{duf12} to estimate  $f(v_e)$\ for the FLAMES-I sample. This is shown in Fig. \ref{v_e} together with that for the VFTS sample \citep{duf12}. They both show a double peaked structure although there are differences in the actual distributions. The number of \vsini\ estimates in the FLAMES-I sample was relatively small (73) and may have some additional selection effects compared with the VFTS sample as discussed in Sect.\ \ref{d_obs}. We have therefore adopted the $f(v_e)$ from the VFTS B-type single star sample in most of our simulations but discuss this choice further below.

\citet{eva15}  classified 434 VFTS targets as B-type and subsequently \citet{dun15} found 141 targets (including 40 supergiants) to be radial velocity variables. \citet{mce14} identified an additional 33 targets as single B-type supergiants, leading to 260 remaining targets. However the relatively low data quality of five of these targets prevented \citet{duf12} estimating projected rotational velocities. Hence we have adopted $N=255$ for the number of B-type VFTS targets which have luminosity classes V to III and have estimated projected rotational velocities. For the FLAMES-I sample, \citet{eva06} has provided spectral types including targets classified as Be-Fe without a luminosity classification. As they are likely to be lower luminosity objects, they were included when deducing a value of $N=103$. Using these sample sizes, we have numerically integrated Equation \ref{e_6} in order to predict the number of nitrogen enriched targets. To validate these results and also to estimate their uncertainties, we have also undertaken a Monte-Carlo simulation. Samples of $N$\ targets were randomly assigned both rotational velocities (using the rotational velocity distribution, $f(v_e)$\ discussed above) and  angles of inclination (using equation \ref{e_1}). The number of targets with different nitrogen enrichments and in different projected rotational bins were then found using the rotational velocity limits in Table \ref{t_v_e}. 

Our simulations are all based on the assumption that the rotational axes of our targets are randomly orientated. As can be seen from Fig.\ \ref{spatial}, approximately half our targets are situated in four different clusters \citep[with different ages, see, for example,][]{wal97, eva11, eva15} with the remainder spread across the field. Such a diversity of locations would argue against them having any axial alignment. However \citet{cor17} identified such alignments in two old Galactic open clusters, although previously \citet{jac10} found no such evidence in two younger Galactic clusters. If the VFTS axes were aligned, the identification by \citet{duf12} of B-type stars with projected rotational velocities, \vsini\,$>$\,400\,\kms\ \citep[which is close to their estimated critical velocities][]{hua10, tow04}, would then imply a value for the inclination axis, $\sin i$\,$\simeq$\,1. In turn this would lead to all our targets having small rotational velocities thereby exacerbating the inconsistencies illustrated in Table \ref{t_N_sim}.

\subsection{Comparison of simulations with numbers of observed nitrogen enriched targets} \label{d_obs}

The numerical integrations and the Monte Carlo simulations led to results that are indistinguishable, providing reassurance that the methodologies were correctly implemented. The Monte-Carlo simulations are summarised in Table \ref{t_N_sim}, their  standard errors being consistent with Poisson statistics which would be expected given the low probability of observing stars at low angles of inclination. For both samples, the simulations appear to significantly underestimate the number of nitrogen enriched targets with \vsini\,$\leq$\,40 \kms\ and possibly overestimate those with 40\,$<$\,\vsini$\,\leq$\,80 \kms.

We have therefore used the Monte-Carlo simulations to find the probabilities  of observing the number of nitrogen enriched stars summarised in Table \ref{t_N_sim}. When the observed number, $n$ was greater than predicted, we calculated the probability that the number observed would be greater than or equal to $n$. Conversely, when the observed number, $n$ was less than predicted, we calculated the probability that the number observed would be less than or equal to $n$. These probabilities are also listed in Table \ref{t_N_sim} and again very similar probabilities would have been found from adopting Poisson statistics. All the differences for the targets with \vsini\,$\leq$\,40 \kms\ are significant at the 5\% level. By contrast for the samples with  40\,$<$\,\vsini$\,\leq$\,80 \kms, none are significant at the 5\% level.

For the VFTS Monte-Carlo simulations, the choice of $N=255$\ leads to 36$\pm$6 and 31$\pm$5\ targets with \vsini\,$\leq$\,40 \kms\ and  40\,$<$\,\vsini$\,\leq$\,80 \kms\ respectively. As expected these are in good agreement with the 37 and 30 targets in the original sample in Table \ref{t_Targets}. However thirteen of these targets were not analysed, reducing the sample summarised in Table \ref{t_Atm_2} to 31 (\vsini\,$\leq$\,40 \kms) and 23 (40\,$<$\,\vsini$\,\leq$\,80 \kms) targets. We have therefore repeated the Monte Carlo simulations  using values of $N$\  (220 and 188 respectively) that reproduce the number of targets actually analysed. These simulations are also summarised in  Table \ref{t_N_sim} and again lead to an overabundance of observed nitrogen enriched targets with  \vsini\,$\leq$\,40 \kms\ that is now significant at 1\% level. By contrast the simulations are consistent with the observations for the cohort with 40\,$<$\,\vsini$\,\leq$\,80 \kms.

Even after allowing for the targets that were not analysed, the discrepancy between the observed number of VFTS targets with \vsini\,$\leq$\,40 \kms\ and the simulations may be larger than predicted. As discussed above, both the conservative current rotational velocities, \ve, adopted in Table \ref{t_v_e} and the decision to follow the stellar evolution to a logarithmic gravity of 3.3 dex will lead to the simulations being effectively upper limits. Additionally, for the VFTS subsample with \vsini\,$\leq$\,40 \kms, there are seven targets, where the upper limit on the nitrogen abundance was consistent with \NtoH$>$7.2 dex and four targets consistent with  \NtoH$>$7.5 dex. If any of these had significant nitrogen enhancements that would further increase the discrepancies for this cohort.

The simulations for the smaller FLAMES-I sample are consistent with those for the VFTS sample. For a sample size of  $N=103$, the number of nitrogen enriched stars with  \vsini\,$\leq$\,40 \kms\  is greater than predicted for both \NtoH$\geq$7.2 dex and \NtoH$\geq$7.5 dex. Indeed the Monte-Carlo simulation implies that this difference is significant at the 1\% level. By contrast the cohort with 40\,$<$\,\vsini$\,\leq$\,80 \kms\ is compatible with the simulations. The re-evaluation of the nitrogen abundance estimates in Sect. \ref{s_N} led to a decrease in the number of targets with \NtoH$>$7.2 dex by four. All these targets  (NGC2004/43, 84, 86 and 91) had  \vsini\,$\leq$\,40 \kms\ and adopting the previous nitrogen abundance estimates would have further increased the discrepancy with the simulations for this cohort. The re-evaluation also moved one target into the \NtoH$>$7.5 dex cohort. However this target  (NGC2004/70) has an estimated \vsini = 46\,\kms\  and hence this change would not affect the low probabilities found for the nitrogen enhanced targets with \vsini\,$\leq$\,40 \kms.

The FLAMES-I results should however be treated with some caution.  Firstly, although there were multi-epoch observations for the this sample, the detection of radial-velocity variables (binaries) was less rigorous than in the VFTS observations; this is discussed further in Sect.\ \ref{d_bin}. Secondly, the exclusion of a number of Be-type stars in the original target selection for NGC\,2004  may have led to this sample being biased to stars with lower values of the rotational velocity  \citep[see][for a detailed discussion of the biases in this sample]{bro11b}. Hence, the use of the VFTS rotational-velocity distribution, $f(v_e)$, may not be appropriate for the FLAMES-I sample. There is some evidence for this in that the observed number of FLAMES-I targets (18 with \vsini\,$\leq$\,40 \kms\ and 16 with 40\,$<$\,\vsini$\,\leq$\,80 \kms) is larger than that predicted (14.5$\pm$3.5 and 12.6$\pm$3.3 respectively). This would be consistent with the $f(v_e)$\ distribution for the FLAMES-I sample shown in Fig.\ \ref{v_e} being weighted more to smaller rotational velocities than the VFTS distribution that was adopted. We have therefore also  used the $f(v_e)$\ distribution, together with the number of \vsini\ estimates, $N=73$ deduced for the FLAMES-I sample in the Monte Carlo simulation and the results are again summarised in Table \ref{t_N_sim}. This leads to even lower  probabilities for the observed number of nitrogen enhanced targets with \vsini\,$\leq$\,40 \kms\,being consistent with the simulations, whilst the probabilities for the 40\,$<$\,\vsini$\,\leq$\,80 \kms\ cohort are increased.\footnote{As the observed number of targets for the 40\,$<$\,\vsini$\,\leq$\,80 \kms\ cohort is now larger than that predicted by the simulations, the probability listed in Table \ref{t_N_sim} is for the observed number of targets (or a greater number) being observed.} The predicted number of targets in the two \vsini\ ranges (15.5$\pm$3.5 and 16.6$\pm$3.6 respectively) are now in better agreement with those observed.

As discussed in Sect.\ \ref{s_Atm}, our adopted \vsini\ estimates may be too high. \citet{sim14} estimated that this might be typically 25$\pm$20\,\kms\ for O-type stars and B-type supergiants; the effects on our samples of B-type dwarfs and giants are likely to be smaller given their smaller macroturbulences \citep{sim17}. However we have considered the effect of arbitrarily decreasing all our \vsini\ estimates by 20\,\kms. This would move eleven VFTS targets and seven FLAMES-I targets into the \vsini\,$\leq$\,40 \kms\ bins. Similar numbers of targets would also be moved into the 40\,$<$\,\vsini$\,\leq$\,80 \kms\ bins from higher projected rotational velocities but as these targets have not been analysed, we are unable to quantitatively investigate the consequences for this bin. However we note that the simulation showed that the predicted number of targets in the 40\,$<$\,\vsini$\,\leq$\,80 \kms\ bin were consistent with that observed and it is likely that this result would remain unchanged.

For the cohorts with \vsini$\leq$\,40 \kms, the numbers would be increased to 48 (VFTS) and 24 (FLAMES-I). However the number of targets with  \NtoH$>$7.2 or 7.5 dex would also be  increased to eight and six (VFTS) and eight and five (FLAMES-I) respectively. We have undertaken Monte-Carlo simulations that reproduce the total number of targets with  \vsini\,$\leq$\,40 \kms\ by increasing the B-type samples to $N$=$335$\ (VFTS) and 170 (FLAMES-I). These simulations again lead to small probabilities of observing the relatively large numbers of nitrogen enhanced targets, viz $<2$\% (VFTS) and  $<0.1$\% (FLAMES-I). This arbitrary decrease in the \vsini\ estimates would also affect the rotational velocity function, $f(v_e)$, used in the simulations. If applied universally, this would shift the distribution to smaller rotational velocities, thereby decreasing the predicted number of nitrogen enhanced targets. Additionally either allowing for  VFTS targets without nitrogen abundance estimates or the use of the FLAMES-I rotational velocity distribution shown in Fig.\ \ref{v_e} would also decrease the predicted numbers and hence probabilities.

Inspection of the metal absorption lines in the spectra for the targets moved into the \vsini\,$\leq$\,40 \kms\ bin showed that they generally had the bell shaped profile characteristic of rotational broadening. Hence we believe that the arbitrary decrease of 20\,\kms\ probably overestimates the magnitude of any systematic errors. As such, we conclude that uncertainties in the \vsini\ estimates are very unlikely to provide an explanation for the discrepancy between the observed and predicted numbers of nitrogen enriched targets in the cohort with \vsini\,$\leq$\,40 \kms.

In summary, the simulations presented in Table \ref{t_N_sim} are consistent with the nitrogen enriched targets for both the VFTS and FLAMES-I samples with 40\,$<$\,\vsini$\,\leq$\,80 \kms\ having large rotational velocities (and small angle of incidence) leading to significant rotational mixing. By contrast, there would appear to be too many nitrogen enriched targets  in both samples with \vsini\,$\leq$\,40 \kms\ to be accounted for by this mechanism. 

\subsection{Frequency of nitrogen enriched, slowly rotating  B-type stars} \label{d_freq}

The simulations presented in Sect.\ \ref{d_obs} imply that there is an excess of targets with \vsini\,$\leq$\,40 \kms\ and enhanced nitrogen abundances. In the VFTS sample, there are two objects with a nitrogen abundance, 7.2$\leq$\NtoH$<$7.5 dex. Our Monte-Carlo simulations imply that there should be 0.85$\pm$0.92 and that the probability of observing two or more objects would be approximately 20\%. Hence their nitrogen enhancement could be due to rotational mixing, although we cannot exclude the possibility that they are indeed slowly rotating. In turn this implies that the excess of targets with high nitrogen abundances appears to occur principally for \NtoH$\geq$7.5 dex. Additionally the small number of such objects in the cohort with 40\,$<$\,\vsini$\,\leq$\,80 \kms\ further implies that these objects have rotational velocities, $v_e\leq 40$\kms. 

In the subsequent discussion, we will consider two percentages, viz. firstly the percentage,  P$_{40}$, of stars {\em within the cohort with $v_e\leq$40 \kms} that have enhanced nitrogen abundances that do not appear to be due to rotational mixing and secondly the percentage, P$_{\mathrm{T}}$, of such stars within the total population of single stars with luminosity classes III-V. Our simulations indicate that 67\% of the 31 VFTS targets with \vsini\,$\leq$\,40 \kms\ will also have $v_e\leq$40 \kms, equating to $\sim$21 targets. From the simulations summarised  in Table \ref{t_N_sim}, the excess of VFTS objects with \NtoH$\geq$7.5 dex and \vsini\,$\leq$\,40 \kms\  is 3.55-3.75 depending on the value of $N$\ adopted. This would then imply  a percentage of such objects,   P$_{40}$$\sim$18\%. This may be an underestimate as there are four targets with $v_e\leq 40$\kms, whose nitrogen abundance cannot be constrained to less than 7.5 dex and excluding these would increase the percentage to P$_{40}$$\sim$21\%. A similar calculation for the FLAMES-I sample leads to a higher fraction of P$_{40}$$\sim$30--33\%  depending on the simulation adopted, with no correction now being required for stars whose nitrogen abundance could not be constrained.

These percentages then translate in to percentages with respect to the whole population B-type stars of P$_{\mathrm{T}}$$\sim$1.5--2.0\% and 3.4--3.7\% for the VFTS and FLAMES-I samples respectively. The latter is significantly smaller than the fraction of slow rotators found by \citet{hun08b} that showed significant nitrogen enrichment, which was estimated as approximately 20\% of the non-binary core hydrogen burning sample. However the two estimates are not directly comparable for the following reasons.

 Firstly \citet{hun08b} adopted a larger projected rotational velocity range, \vsini\,$\leq$\,50 \kms, to that adopted here. Secondly  they defined a significant nitrogen enrichment as \NtoH$\geq$7.2 dex rather than the larger value adopted here. Thirdly they selected all targets with estimated effective temperatures  \teff$\leq$\,35\,000\,K and gravities \logg$\geq$\,3.2 dex. These included some late O-type stars, supergiants and binaries. Fourthly they made no allowance for the targets with  \,$\leq$\,50 \kms\ but with a rotational velocity sufficient to produce the observed nitrogen abundance by rotational mixing. Finally they adopted for their LMC non-binary core hydrogen burning sample the 73 apparently single targets that had estimates for their projected rotational velocity. However as discussed in Sect. \ref{d_obs}, \citet{eva05} lists 103 B-type non-supergiant stars with no evidence for binarity and we have adopted this larger sample size in our simulations.

 \citet{hun08b} identified 17 targets (three of which were binaries) that had nitrogen enhancements and small projected rotational velocities.  Of the single targets two targets had spectral types of O9.5 III or O9.5 V, whilst an additional target was a B0.5 Ia supergiant. Removing these six objects (in order to maintain consistency with the current analysis) would reduced the sample size to 11 objects.  A simulation similar to those described in Sect. \ref{d_sim} implied that 1.48$\pm$1.21 of these targets were consistent with rotational mixing. Adopting a sample size, $N$=103  would then lead to a percentage, P$_{\mathrm{T}}$$\sim$9\% of FLAMES-I B-type targets with luminosity classes III-V, \vsini$\leq$50 \kms, \NtoH$\geq$7.2 dex that could not be explained by rotational mixing.  We note that this revised estimate is consistent with that discussed below for the VFTS dataset when adopting similar criteria. 

By contrast our current estimate of P$_{\mathrm{T}}$\,$\sim$\,2-4\%  for the B-type targets with luminosity classes III-V, \vsini$\leq$40 \kms\ and  \NtoH$\geq$7.5 dex. should be considered as strong lower bound to the actual fractions of slowly rotating targets that can not be explained by currently available single star evolutionary models. This arises from both the conservative assumptions that we have adopted (see Sect. \ref{d_sim} and \ref{d_obs}) and the possibility that there may be additional  targets with smaller nitrogen enrichments that are inconsistent with the predictions of  rotational mixing. 

For example, the VFTS sample has a median logarithmic gravity of 3.8 dex and median initial mass (see Sect.\ \ref{d_bonnsai}) of 11.2\Msun. The models of \citet{bro11a} with an initial mass of 12\Msun\ then  imply that to obtain a nitrogen enhancement of 0.2 dex as a star evolves to this gravity would require a current rotational velocity of 150\,\kms. We note that this velocity is not directly comparable with those in Table\ \ref{t_v_e} as we are now considering median gravities rather than those found at the end of the hydrogen core-burning phase. 

The number of targets in our sample of $N=255$\ predicted by Monte Carlo simulations would then be 2.16$\pm$1.45 compared with the eight targets that are observed (see Table \ref{t_Atm_2}) leading to  percentages of P$_{\mathrm{T}}$$\sim$ 2.3\% and P$_{40}$$\sim$28\%. Including those targets whose nitrogen abundances are not constrained to less than 7.1 dex would increase these percentages to  P$_{\mathrm{T}}\la 6.2$\% and P$_{40}\la 76$\%. A similar calculations for the FLAMES-I sample (with $N$=103) would predict 0.87$\pm$0.93 nitrogen enriched targets compared with the nine stars (see Table \ref{t_FLAMESI}) actually observed. This would then translate to percentages of P$_{\mathrm{T}}$$\sim$7.9\% and P$_{40}$$\sim$71\%.  For this sample, there are no stars with unconstrained nitrogen abundances. We stress that these estimates should be considered illustrative as, for example, the nitrogen enrichment of 0.2 dex is similar to the expected errors. 

In summary the percentage of targets with \ve$\leq$40\,\kms and \NtoH$\geq$7.5 dex in the VFTS and FLAMES-I sample are of the order of P$_{\mathrm{T}}$$\sim$2--4\% and P$_{40}$$\sim$20--30\%. Given the large nitrogen enhancement and conservative assumptions that have been adopted, these constitute strong lower limits to the number of slowly rotating stars that are inconsistent with current single star evolutionary grids of models. Considering lower nitrogen enhancements and using evolutionary predictions for the median gravities of our models implies that the frequency of such stars could be substantially larger, viz. P$_{\mathrm{T}}\sim$6--8\% and  P$_{40}\sim$70\%.

\subsection{Origin of the nitrogen enriched targets} \label{d_origin}

Below we discuss possible explanations for the origin of the targets discussed in Sect.\  \ref{d_freq} -- hereafter referred to as 'nitrogen enhanced'.  

\subsubsection{Comparison with the predictions of single star evolutionary models} \label{d_bonnsai}

The estimated atmospheric parameters for the VFTS and FLAMES-I samples are shown in Fig. \ref{atm_par}, together with the predictions of the evolutionary models of \cite{bro11a}.  The latter are for effectively zero initial rotational velocity and were chosen to be consistent with the observed low projected rotational velocities of our targets. However, in this part of the HR diagram, the evolutionary tracks and isochrones are relatively insensitive to the choice of initial rotational velocity as can be seen from Figs. 5 and 7 of \citet{bro11a}.

 Also shown in Fig. \ref{atm_par} are the Geneva tracks and isochrones for an LMC metallicity and zero rotational velocity extracted or interpolated from the Geneva stellar models database\footnote{The Geneva database is available at \newline \url{https://obswww.unige.ch/Recherche/evol/}} for the grid of models discussed by \citet{geo13a}. These models had a maximum initial mass of 15\,\Msun, so no track for 19\,\Msun\ is shown, whilst the lower age isochrones are also truncated. As can be seen from Fig.\ \ref{atm_par}, the two sets of tracks and isochrones are in reasonable agreement but show some differences. In particular the end of the terminal age main sequence occurs at different positions on the two sets of tracks. These differences have previously been discussed by \citet{cas14}, who ascribed them to the different calibrations of the convective core overshooting parameter. However in the context of the current analysis, the two grids would lead to similar estimates of, for example, stellar masses and ages within the estimated uncertainties in the atmospheric parameters. For both samples, there is some evidence that the nitrogen enhanced targets may lie in the higher \teff regions of the diagrams. However Student t-tests and Mann-Whitney U-tests returned no statistics that were significant at the 10\% level for their being any differences in the effective temperatures distributions of the parent populations. 
 
 \citet{bro11b} have previously discussed the masses of the B-type FLAMES-I targets based on the analyses of \citet{hun07, hun08b, hun09a} and \citet{tru07}. They noted that the nitrogen enhanced slowly spinning stars appeared to have higher masses. More recently \citet{gri17} have estimated nitrogen abundances for O-type giants and supergiants from the VFTS and again find that the targets with small \vsini\ and enhanced nitrogen abundance have higher estimated masses than the rest of their sample. However it should be noted that these two analyses sample different mass ranges with the B-type stars having estimated masses in the range $\sim$8-15\Msun\ (see Fig.\ \ref{atm_par}), whilst those of \citet{gri17}  normally have estimated masses $>20$\Msun.

Following \citet{gri17}, we have used {\sc Bonnsai} \footnote{The {\sc Bonnsai} web-service is available at \newline \url{www.astro.uni-bonn.de/stars/bonnsai}.} to estimate the initial and current evolutionary masses of the stars in our VFTS and FLAMES-I samples with \vsini\,$\leq$\,40\,\kms.  {\sc Bonnsai} uses a Bayesian methodology and the grids of models of \citet{bro11a} to constrain the evolutionary status of a given star, including its age and mass  \citep[see][for details]{sch14}. As independent prior functions, we adopted the LMC metallicity grid of models, a \citet{sal55} initial mass function, the initial rotational velocity distribution estimated by \citet{duf12}, a random orientiation of spin axes, and a uniform age distribution. The estimates of \teff and \logg\ (taken from Tables \ref{t_Atm_2} and \ref{t_N}), together with a \vsini$\le$40\,\kms\ were then used to estimate masses. For these  relatively unevolved stars, the predicted current and initial masses were very similar with differences $<5$\%, apart from one VFTS target, VFTS\,183, where the difference was 8\%. 

For both samples the nitrogen enriched stars had higher mean masses, confirming the previous comments for the FLAMES-I sample by \citet{bro11b}. For the VFTS sample, the nitrogen enriched targets had a median current mass of 14\,\Msun\ and a mean current mass of 13.6$\pm$2.1\,\Msun\ compared with 11.2 and 11.8$\pm$2.9\,\Msun\ for the other stars. For the FLAMES-I sample,  the  corresponding values  were 12.5 and 12.7$\pm$1.4\,\Msun\ compared with 10.4 and 11.8$\pm$2.4\,\Msun. However these differences are relatively small especially given the sample sizes for the nitrogen enriched objects. This was confirmed by Student t-tests and Mann-Whitney U-tests that returned no statistics that the two parent populations were different at the 5\% level. 

The {\sc Bonnsai} analysis also provided estimates for the stellar luminosity, lifetime and current surface nitrogen abundance. There is good agreement between the {\sc Bonnsai} luminosities and the observed luminosities estimated in Sect. \ref{s_obs} and listed in Table \ref{t_Targets}. The mean difference is $-0.02\pm 0.22$ dex with the estimates for over $\sim$85\% of the targets showing a difference of less than 0.3 dex. The major discrepancy is for VFTS\,469, where {\sc Bonnsai} estimates a luminosity 0.6 dex (i.e. a factor of four) greater than the observed luminosity. As discussed in Sect.\ \ref{s_obs}, the spectroscopy of this star may be compromised by contamination by light from a bright O-type star in an adjacent fibre.  Excluding this star would lead to a mean difference of  $0.00\pm 0.19$ dex. For the five nitrogen enhanced stars, the agreement between the two sets of luminosities is better, with the mean difference being $-0.02\pm 0.06$ dex.

\begin{figure}
\epsfig{file=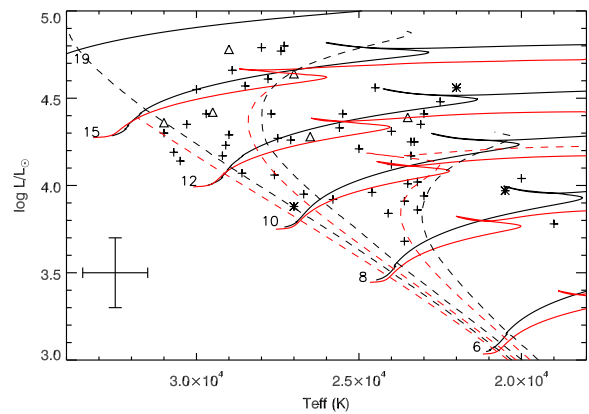,width=9cm, angle=0} 
\caption{Observed luminosities for the VFTS sample as a function of effective temperature. Targets with an estimated nitrogen abundance, 7.2$\le$\NtoH$<$7.5 dex, are  shown as asterisks and those with \NtoH$\ge$7.5 dex  are shown as triangles; all other targets are shown as crosses. Evolutionary tracks and isochrones (see caption to Fig.\ \ref{atm_par}) are also shown, together with typical uncertainties in the effective temperatures and luminosity estimates.}
\label{l_teff}
\end{figure}

In Fig. \ref{l_teff}, the observed luminosities estimated in Sect.\ \ref{s_obs} are shown as a function of effective temperature and imply that the nitrogen enhanced stars may have larger luminosities. Indeed these stars have both higher mean observed (4.45$\pm$0.19 dex) and  {\sc Bonnsai} (4.47$\pm$0.22 dex)  luminosities than the other targets (4.20$\pm$0.27 dex  and 4.42$\pm$0.27 dex  respectively). This is consistent with the differences in the masses discussed above. Student t-tests and Mann-Whitney U-tests returned probabilities that the two parent populations were the same at the $\sim$5-6\% level apart from the t-test for the observed luminosities where the probability was $\sim$3\%.

The lifetimes estimated by {\sc Bonnsai} ranged from $\sim$0--30\,Myr with there being significant uncertainties in individual values due to the samples lying close to the zero age main sequence and the age estimates thereby being very sensitive to the adopted gravity. The FLAMES-I sample have on average larger age estimates than those for the VFTS sample, which is consistent with the target selection \citep[see][for details]{eva06,eva11}. The median and average of the predicted ages of the VFTS nitrogen enriched targets (8.1 and 9.5$\pm$4.4 Myr) were lower than those for the other stars (10.9 and 12.6$\pm$7.7 Myr). Similar differences were found for the FLAMES-I sample, viz. nitrogen enhanced stars: 9.3 and 9.0$\pm$4.42\ Myr; other stars: 14.0 and 14.3$\pm$8.8 Myr. However again statistical tests showed that these differences were not significant at a 5\% level.

The nitrogen abundances predicted by {\sc Bonnsai} ranged from 6.89--6.91\,dex  -- in effect implying no nitrogen enhancement from the initial value of 6.89\,dex adopted in the models of \citet{bro11a}. Additionally the  predicted initial and current rotational velocities were small and less than or equal to 40\,\kms. This is consistent with the small \vsini\ ($\le$40\,\kms) estimates of these samples and the low probability of observing targets at small angles of incidence. Indeed it qualitatively re-enforces the results presented in Sect.\ \ref{d_obs}, which found that the numbers of observed nitrogen enhanced targets were significantly larger than predicted by our simulations.

For the nitrogen enhanced objects, {\sc Bonnsai} was re-run including the estimated nitrogen abundance as a further constraint for all nine (five VFTS and four FLAMES-I)  targets. In all cases, the best fitting model had both high initial (\vi) and current (\ve) rotational velocities, which were consistent with the adopted limits given in Table \ref{t_v_e} to achieve significant nitrogen enhancements by rotational mixing. However for four targets (VFTS\,095 and 881, NGC2004/53 and N11/100), the {\sc Bonnsai} solutions failed both the posterior predictive check and a\ $\rchi^2$-test. This difficulty in finding single star evolutionary models that match the observed parameters of our nitrogen enhanced targets again re-enforces the inconsistencies discussed in Sect.\ \ref{d_sim}.

To summarise, there is some evidence that the nitrogen enriched targets with \vsini\,$\le$\,40\,\kms\ may have higher masses, effective temperatures and luminosities, together with smaller lifetimes, assuming that they have evolved as single stars. The higher mass estimates are consistent with a previous population synthesis for the FLAMES-I B-type stars by \citet{bro11b} and the analysis of the evolved VFTS O-type targets by \citet{gri17}. 

%---------------------------------------------------------------------------------------------
\subsubsection{Physical assumptions in single star evolutionary models} \label{d_phy}

The discrepancy between the observed and predicted numbers of nitrogen enhanced targets discussed in Sect. \ref{d_obs} could be due to the physical assumptions adopted in the stellar evolutionary models. For example,  as discussed by \citet{gri17}, the efficiency of rotational mixing in the single-star evolutionary models of \citet{bro11a} was calibrated using efficiency factors based on the B-type dwarfs in the FLAMES-I survey but excluding nitrogen enhanced targets with low projected rotational velocities. Hence an increase in mixing efficiencies could in principle explain our nitrogen enriched targets. For example the five VFTS targets with nitrogen enhancements of greater than 0.6\,dex  and \vsini$\le 40$\,\kms\ (see Table \ref{t_N_sim}) could be reproduced by reducing the current rotational velocity, \ve\ for such an enhancement, to the unrealistically low value of $\sim$\,70\,\kms. However this would also increase the number of nitrogen enriched targets in the  40\,$<$\,\vsini$\,\leq$\,80 \kms\ cohort to  $\sim$\,20\ far higher than is observed. Similar arguments apply to the FLAMES-I sample where the fraction of nitrogen enhanced targets is even larger. Indeed the ratios of nitrogen enriched targets in the two projected rotational velocity cohorts provides strong evidence against the observed nitrogen enriched stars being rotationally mixed stars observed at small angles of inclination.

Another possibility is our nitrogen enriched targets may have had severe stripping of their envelopes, thereby revealing chemically enriched layers. The associated mass loss could also have removed angular momentum from the surface, resulting in spin-down. However the predicted mass loss for our targets from the {\sc Bonnsai} simulations or from the grid of models of \citet{geo13a} is very small (typically less than 5\% of the initial mass) making such an explanation implausible.

In summary, there is no evidence that limitations on the physical assumptions made in the stellar evolutionary models can explain our nitrogen enhanced targets. Indeed the similarity of their physical properties to those of the other targets (discussed in Sect.\ \ref{d_bonnsai}) argues against such an explanation.

\subsubsection{Magnetic fields in single stars}

The models discussed in Sect. \ref{d_bonnsai} and \ref{d_phy} assume that magnetic fields are unimportant for a star's evolution. Such fields have been previously  suggested as an explanation for nitrogen enriched, slowly rotating Galactic early-B stars \citep{mor08,mor12,prz11}.  However, more recently,  \citet{aer14a} found no significant correlation between nitrogen abundances and magnetic field strengths in a statistical analysis of 68 Galactic O-type stars that addressed the incomplete and truncated nature of the observational data.

\citet{mey11} discussed the possibility of  magnetic braking \citep[see][and references therein]{udd09} during the main-sequence phase. Models with differential rotation and magnetic braking produced strongly mixed stars with low surface rotational velocities. By contrast models with solid-body rotation and magnetic braking produce stars that at the terminal age main sequence had low surface rotational velocities and effectively no changes in the surface abundances. \citet{pot12b, pot12a} have considered the $\alpha-\Omega$ dynamo as a mechanism for driving the generation of large-scale magnetic flux.  They found that this mechanism was effective for stellar masses up to $\sim$15 \Msun\ (which includes effectively all the stars in our samples). B-type main sequence stars with sufficiently high rotation rates were found to develop an active dynamo and so exhibited strong magnetic fields. They then were spun down quickly by magnetic braking and magneto-rotational turbulence \citep{spr02} leading to enhanced surface nitrogen abundances. 

Both observational and theoretical  investigations have therefore led to inconclusive results about whether magnetic fields could enhance stellar mixing. However it is an attractive mechanism for explaining our nitrogen enhanced stars as it can combine both high nitrogen abundances and low stellar rotation velocities. The incidence of magnetic fields in early-type stars is currently being investigated by the MiMes survey \citep{wad14, wad16} and by the BOB survey \citep{mor14, mor15}. The MiMes survey had a number of criteria for selecting their survey sample, one of which favoured stars with \vsini\,$\leq$\,150\,\kms. They found a frequency of  7$\pm$1\% for magnetic fields in 430 B-type stars \citep{wad14} with a similar frequency for O-type stars \citep{gru17}.  The BOB survey have found a detection rate of 6$\pm$4\% for a sample of 50 OB-type stars, while by considering only the apparently slow rotators they derived a detection rate of 8$\pm$5\%  \citep{fos15a}. Additional observations for 28 targets led to revised estimates of 6$\pm$3\% and 5$\pm$5\% \citep{sch17}.

These detection frequencies appear to be consistent with the percentages of nitrogen enriched targets estimated for the VFTS and FLAMES-I surveys (see Sect. \ref{d_freq}). However the two magnetic surveys have necessarily only observed Galactic targets and may not be directly applicable to an LMC metallicity. Also both surveys preferentially detect targets with large magnetic fields, typically $>$100\,Gauss \citep{mor15}. \citet{fos15b} on the basis of intensive observations of two very bright stars $\beta$ CMa and $\epsilon$ CMa  detected relatively weak magnetic fields and concluded that such fields might be more common in massive stars than currently observed. However the simulations of \citet{mey11} for these magnetic field strengths imply that the enhancement of nitrogen occurs slowly during the hydrogen burning phase. In turn this would imply that such stars might appear to be older, which is not observed with our nitrogen enhanced targets.

\citet{fos16} have estimated the fundamental parameters including ages, for a sample of 389 early-type stars,  including 61 stars that have magnetic fields. The atmospheric parameters of the latter appear similar to those of the other hydrogen burning main sequence (MS) stars and they concluded that `the fraction of magnetic massive stars remains constant up to a fractional MS age of $\simeq$0.6 and decreases rapidly at older ages'. Hence there is no evidence for magnetic stars being older, although the rate of nitrogen enrichment would appear to depend on their magnetic field strength. 

Stars with magnetic fields could account for our observed nitrogen enhanced stars and indeed this would appear to be a promising explanation. However, it should be noted that the lack of nitrogen enhanced primaries in the VFTS binary systems \citep{gar17} and the existence of nitrogen enhanced targets amongst the O-type stars \citep[see][and references therein]{gri17}, where the mechanism proposed by \citet{pot12a} would not appear to be applicable,  imply that more than one mechanism might be responsible.

\subsubsection{Extant binary systems} \label{d_bin}

The analysis of \citet{dun15} of the VFTS B-type sample identified targets that were members of a binary systems, both SB1 and SB2. However the remaining targets (including those with low projected rotational velocities considered here) will contain undetected binary systems. \citet{dun15} estimated that that whilst their observed binary fraction was 0.25$\pm$0.02, the true binary fraction was 0.58$\pm$0.11. In turn this implies that approximately half of the sample may in fact be binary system. These will be weighted towards long period system with periods,  $P>100$ days, where the probability of detection decreases; these pre-interaction systems might be expected to evolve as single stars. However there will also be shorter period systems (often with small orbital angles of inclination) that remain undetected and these could possibly explain the nitrogen enriched objects.

\citet{gar17} analysed VFTS spectroscopy to estimate the atmospheric parameters of the primaries with \vsini\,$<80$\,\kms\ in 33 B-type binaries. Surprisingly none of these primaries had \NtoH$\geq$7.5 dex and in only one case was the upper limit consistent with such a nitrogen abundance. Indeed a preliminary comparison by  \citet{gar17} of their results with those found here implied that their binary sample did not contain significantly nitrogen enriched binaries. They concluded that their systems were in a pre-interaction epoch and evolving as single stars with low rotational velocities and small amounts of rotational mixing \citep{deM09}. Hence it would appear unlikely that the nitrogen enrichments in our sample have been produced by processes in targets that are {\em currently} pre-interaction binaries. The simulation of \citet{dun15} would then imply that if our nitrogen enriched targets are indeed single,  they may make up an even larger fraction of the single B-type star cohort with $v_e\leq 40$\,\kms; the actual increase would depend on the frequency of undiscovered binaries with different projected rotational velocities.

\subsubsection{Post-interaction binaries}

Another possible explanation of the nitrogen enhanced targets is that they are the product of binary systems in which the components have interacted. \citet{deM14} discussed how such interactions could increase the fraction of  stars that were classified as single on the basis of a lack of radial velocity variations. Firstly in the case of a stellar merger, the product would now be a single star. Secondly, following mass transfer, the mass gainer would dominate the light from the system (precluding an SB2 classification), whilst typically only having modest radial velocity variations (less than 20\,\kms). 

\citet{ram13} suggested that such a population of post-interaction binaries may explain the presence of a high-velocity tail (\vsini\,$> 300$\,\kms) found in the rotational velocity distribution of the VFTS `single' O-type star sample, which had been initially proposed by \citet{deM13}. Further support for this was the apparent absence of such a high rotational velocity tail for the pre-interacting VFTS O-type binaries analysed by \citet{ram15}.

The possibility that the nitrogen enhanced FLAMES-I targets might be  binary products was considered by \citet{bro11b}.  For example, as discussed by \citet{lan08}, post-mass transfer systems may remain so tightly bound that tidal interaction could spin down the (observed) accreting component. However \citet{bro11a} concluded that  `close binary evolution, as far as it is currently understood, is unlikely to be responsible for the slowly rotating nitrogen-rich population of observed stars'. This was based on the binary star pathways discussed by \citet{lan08} implying that most post-interaction objects would be rapidly rotating.

Magnetic fields could also play a role in binary systems via mergers of pre-main or main sequence stars \citep{sch16}.  The former could occur via tidal interactions with circumstellar material \citep{sta10, kor12}, whilst the latter would occur in some binary evolutionary pathways \citep{pod92, lan12}.  Then the nitrogen enrichment would be due to the merger and its aftermath and/or rotational mixing of the rapidly rotating product. If the stellar merger also produced a magnetic field, this could then be an efficiently spin down mechanism. Products of stellar mergers may represent $\sim$10\% of the O-type field population (de Mink et al. 2014). If mergers represented a similar fraction of the B-type population, this would be consistent with the percentages of nitrogen enriched stars discussed in Sect. \ref{d_freq}. Additionally the products of such mergers might be expected to be more massive and appear younger than other single B-type stars. Tentative evidence for this can be seen in Fig. \ref{atm_par} and \ref{l_teff} and has been discussed in Sect.\ \ref{d_bonnsai}.

In summary, post-interaction binary evolutionary pathways are available that might produce slowly rotating, nitrogen enriched, apparently single stars, either by mass transfer or mergers. The latter would also be consistent with the masses and ages inferred for our stellar samples. However models exploring the wide variety of possible evolutionary pathways are required for further progress to be made.

\section{Conclusions}

Our principle conclusions are that:

\begin{enumerate}
\item Approximately 75\% of the targets in the FLAMES-I and VFTS surveys with \vsini\,$\leq$\,80\,\kms\ have nitrogen enhancements of less than 0.3 dex. As such they would appear to be slowly rotating stars that have not undergone significant mixing.
\item Both surveys contain stars that exhibit significant nitrogen enhancements. The relative numbers in the cohorts with projected rotational velocities, \vsini\ of 0--40\,\kms\ and 40--80\,\kms\ are inconsistent with these being rapidly rotating stars observed at small angles of inclination.
\item For both surveys, the number of targets with  \vsini\,$\leq$\,40\,\kms\  and \NtoH\,$\geq$\,7.5 dex is larger than predicted. These differences are significant at a high level of probability.
\item The percentage of these highly nitrogen enhanced objects that cannot be explained by rotational mixing are estimated to be $\sim$20-30\% of single stars with current rotational velocities of less than 40\,\kms and $\sim$2-4\% of the total non-supergiant  single B-type population. Given the conservative assumptions adopted in the simulations and the large nitrogen enhancements observed, these constitute robust {\em  lower limits} for stars that appear inconsistent with current grids of single star evolutionary models incorporating rotational mixing.
\item Including targets with smaller nitrogen enhancements and adopting evolutionary models consistent with the median gravities and masses of our samples leads to larger estimates of the fractions of targets that are inconsistent with current evolutionary models, viz.  $\sim$70\%  with current rotational velocities less than 40\,\kms\ and $\sim$6-8\% of the total single B-type stellar population.
\item The highly nitrogen enriched targets may have higher effective temperatures, masses and luminosities than the other stars.  Additionally single star evolutionary models imply that they would be younger, implying that they may be re-juvenated. However these differences are generally not significant at the 5\% level.
\item Some of our targets will be undetected binaries as discussed by \citet{dun15} and this binary population does not appear to contain highly nitrogen enhanced targets \citep{gar17}. Hence the percentages of {\em truly single} stars that have significant nitrogen abundances may be higher than has been estimated. Indeed it is possible that effectively all the single B-type population with a current rotational velocity, \ve$\leq$40\,\kms\ may have nitrogen abundances that are  inconsistent with single star evolutionary models.
\item Possible explanations for these nitrogen enhancements are considered of which the most promising would appear to be breaking due to magnetic fields or stellar mergers with subsequent magnetic braking. The latter would be consistent with the higher masses that may pertain in the nitrogen enriched sample.
\end{enumerate}

\begin{acknowledgements}
Based on observations at the European Southern Observatory Very Large Telescope in programme 182.D-0222. SdM acknowledges support by a Marie Sklodowska-Curie Action (H2020 MSCA-IF-2014, project id 661502). 
\end{acknowledgements}

\bibliography{literature.bib}

\newcommand{\noop}[1]{}
\begin{thebibliography}{129}
\expandafter\ifx\csname natexlab\endcsname\relax\def\natexlab#1{#1}\fi

\bibitem[{{Abt} {et~al.}(2002){Abt}, {Levato}, \& {Grosso}}]{abt02}
{Abt}, H.~A., {Levato}, H., \& {Grosso}, M. 2002, \apj, 573, 359

\bibitem[{{Aerts} {et~al.}(2014){Aerts}, {Molenberghs}, {Kenward}, \&
  {Neiner}}]{aer14a}
{Aerts}, C., {Molenberghs}, G., {Kenward}, M.~G., \& {Neiner}, C. 2014, \apj,
  781, 88

\bibitem[{{Ahmed} \& {Sigut}(2017)}]{ahm17}
{Ahmed}, A. \& {Sigut}, T.~A.~A. 2017, \mnras, 471, 3398

\bibitem[{{Allende Prieto} {et~al.}(2003){Allende Prieto}, {Lambert}, {Hubeny},
  \& {Lanz}}]{all03}
{Allende Prieto}, C., {Lambert}, D.~L., {Hubeny}, I., \& {Lanz}, T. 2003,
  \apjs, 147, 363

\bibitem[{{Asplund} {et~al.}(2009){Asplund}, {Grevesse}, {Sauval}, \&
  {Scott}}]{asp09}
{Asplund}, M., {Grevesse}, N., {Sauval}, A.~J., \& {Scott}, P. 2009, \araa, 47,
  481

\bibitem[{{Brott} {et~al.}(2011{\natexlab{a}}){Brott}, {de Mink}, {Cantiello},
  {Langer}, {de Koter}, {Evans}, {Hunter}, {Trundle}, \& {Vink}}]{bro11a}
{Brott}, I., {de Mink}, S.~E., {Cantiello}, M., {et~al.} 2011{\natexlab{a}},
  \aap, 530, A115

\bibitem[{{Brott} {et~al.}(2011{\natexlab{b}}){Brott}, {Evans}, {Hunter}, {de
  Koter}, {Langer}, {Dufton}, {Cantiello}, {Trundle}, {Lennon}, {de Mink},
  {Yoon}, \& {Anders}}]{bro11b}
{Brott}, I., {Evans}, C.~J., {Hunter}, I., {et~al.} 2011{\natexlab{b}}, \aap,
  530, A116

\bibitem[{{Carrera} {et~al.}(2008){Carrera}, {Gallart}, {Hardy}, {Aparicio}, \&
  {Zinn}}]{car08}
{Carrera}, R., {Gallart}, C., {Hardy}, E., {Aparicio}, A., \& {Zinn}, R. 2008,
  \aj, 135, 836

\bibitem[{{Carroll}(1933)}]{car33}
{Carroll}, J.~A. 1933, \mnras, 93, 478

\bibitem[{{Castro} {et~al.}(2014){Castro}, {Fossati}, {Langer},
  {Sim{\'o}n-D{\'{\i}}az}, {Schneider}, \& {Izzard}}]{cas14}
{Castro}, N., {Fossati}, L., {Langer}, N., {et~al.} 2014, \aap, 570, L13

\bibitem[{{Chiosi} \& {Maeder}(1986)}]{chi86}
{Chiosi}, C. \& {Maeder}, A. 1986, \araa, 24, 329

\bibitem[{{Conti} \& {Ebbets}(1977)}]{con77}
{Conti}, P.~S. \& {Ebbets}, D. 1977, \apj, 213, 438

\bibitem[{{Corsaro} {et~al.}(2017){Corsaro}, {Lee}, {Garc{\'{\i}}a},
  {Hennebelle}, {Mathur}, {Beck}, {Mathis}, {Stello}, \& {Bouvier}}]{cor17}
{Corsaro}, E., {Lee}, Y.-N., {Garc{\'{\i}}a}, R.~A., {et~al.} 2017, Nature
  Astronomy, 1, 0064

\bibitem[{{Cranmer}(2005)}]{cra05}
{Cranmer}, S.~R. 2005, \apj, 634, 585

\bibitem[{{de Mink} {et~al.}(2009){de Mink}, {Cantiello}, {Langer}, {Pols},
  {Brott}, \& {Yoon}}]{deM09}
{de Mink}, S.~E., {Cantiello}, M., {Langer}, N., {et~al.} 2009, \aap, 497, 243

\bibitem[{{de Mink} {et~al.}(2013){de Mink}, {Langer}, {Izzard}, {Sana}, \& {de
  Koter}}]{deM13}
{de Mink}, S.~E., {Langer}, N., {Izzard}, R.~G., {Sana}, H., \& {de Koter}, A.
  2013, \apj, 764, 166

\bibitem[{{de Mink} {et~al.}(2014){de Mink}, {Sana}, {Langer}, {Izzard}, \&
  {Schneider}}]{deM14}
{de Mink}, S.~E., {Sana}, H., {Langer}, N., {Izzard}, R.~G., \& {Schneider},
  F.~R.~N. 2014, \apj, 782, 7

\bibitem[{{Donati} \& {Landstreet}(2009)}]{don09}
{Donati}, J.-F. \& {Landstreet}, J.~D. 2009, \araa, 47, 333

\bibitem[{{Doran} {et~al.}(2013){Doran}, {Crowther}, {de Koter}, {Evans},
  {McEvoy}, {Walborn}, {Bastian}, {Bestenlehner}, {Gr{\"a}fener}, {Herrero},
  {K{\"o}hler}, {Ma{\'{\i}}z Apell{\'a}niz}, {Najarro}, {Puls}, {Sana},
  {Schneider}, {Taylor}, {van Loon}, \& {Vink}}]{dor13}
{Doran}, E.~I., {Crowther}, P.~A., {de Koter}, A., {et~al.} 2013, \aap, 558,
  A134

\bibitem[{{Dufton} {et~al.}(2013){Dufton}, {Langer}, {Dunstall}, {Evans},
  {Brott}, {de Mink}, {Howarth}, {Kennedy}, {McEvoy}, {Potter},
  {Ram{\'{\i}}rez-Agudelo}, {Sana}, {Sim{\'o}n-D{\'{\i}}az}, {Taylor}, \&
  {Vink}}]{duf12}
{Dufton}, P.~L., {Langer}, N., {Dunstall}, P.~R., {et~al.} 2013, \aap, 550,
  A109

\bibitem[{{Dufton} {et~al.}(2005){Dufton}, {Ryans}, {Trundle}, {Lennon},
  {Hubeny}, {Lanz}, \& {Allende Prieto}}]{duf05}
{Dufton}, P.~L., {Ryans}, R.~S.~I., {Trundle}, C., {et~al.} 2005, \aap, 434,
  1125

\bibitem[{{Dunstall} {et~al.}(2011){Dunstall}, {Brott}, {Dufton}, {Lennon},
  {Evans}, {Smartt}, \& {Hunter}}]{dun11}
{Dunstall}, P.~R., {Brott}, I., {Dufton}, P.~L., {et~al.} 2011, \aap, 536, A65

\bibitem[{{Dunstall} {et~al.}(2015){Dunstall}, {Dufton}, {Sana}, {Evans},
  {Howarth}, {Sim{\'o}n-D{\'{\i}}az}, {de Mink}, {Langer}, {Ma{\'{\i}}z
  Apell{\'a}niz}, \& {Taylor}}]{dun15}
{Dunstall}, P.~R., {Dufton}, P.~L., {Sana}, H., {et~al.} 2015, \aap, 580, A93

\bibitem[{{Ekstr{\"o}m} {et~al.}(2012){Ekstr{\"o}m}, {Georgy}, {Eggenberger},
  {Meynet}, {Mowlavi}, {Wyttenbach}, {Granada}, {Decressin}, {Hirschi},
  {Frischknecht}, {Charbonnel}, \& {Maeder}}]{eka12}
{Ekstr{\"o}m}, S., {Georgy}, C., {Eggenberger}, P., {et~al.} 2012, \aap, 537,
  A146

\bibitem[{{Evans} {et~al.}(2015){Evans}, {Kennedy}, {Dufton}, {Howarth},
  {Walborn}, {Markova}, {Clark}, {de Mink}, {de Koter}, {Dunstall},
  {H{\'e}nault-Brunet}, {Ma{\'{\i}}z Apell{\'a}niz}, {McEvoy}, {Sana},
  {Sim{\'o}n-D{\'{\i}}az}, {Taylor}, \& {Vink}}]{eva15}
{Evans}, C.~J., {Kennedy}, M.~B., {Dufton}, P.~L., {et~al.} 2015, \aap, 574,
  A13

\bibitem[{{Evans} {et~al.}(2006){Evans}, {Lennon}, {Smartt}, \&
  {Trundle}}]{eva06}
{Evans}, C.~J., {Lennon}, D.~J., {Smartt}, S.~J., \& {Trundle}, C. 2006, \aap,
  456, 623

\bibitem[{{Evans} {et~al.}(2005){Evans}, {Smartt}, {Lee}, {Lennon}, {Kaufer},
  {Dufton}, {Trundle}, {Herrero}, {Sim{\'o}n-D{\'{\i}}az}, {de Koter},
  {Hamann}, {Hendry}, {Hunter}, {Irwin}, {Korn}, {Kudritzki}, {Langer},
  {Mokiem}, {Najarro}, {Pauldrach}, {Przybilla}, {Puls}, {Ryans}, {Urbaneja},
  {Venn}, \& {Villamariz}}]{eva05}
{Evans}, C.~J., {Smartt}, S.~J., {Lee}, J.-K., {et~al.} 2005, \aap, 437, 467

\bibitem[{{Evans} {et~al.}(2011){Evans}, {Taylor}, {H{\'e}nault-Brunet},
  {Sana}, {de Koter}, {Sim{\'o}n-D{\'{\i}}az}, {Carraro}, {Bagnoli}, {Bastian},
  {Bestenlehner}, {Bonanos}, {Bressert}, {Brott}, {Campbell}, {Cantiello},
  {Clark}, {Costa}, {Crowther}, {de Mink}, {Doran}, {Dufton}, {Dunstall},
  {Friedrich}, {Garcia}, {Gieles}, {Gr{\"a}fener}, {Herrero}, {Howarth},
  {Izzard}, {Langer}, {Lennon}, {Ma{\'{\i}}z Apell{\'a}niz}, {Markova},
  {Najarro}, {Puls}, {Ramirez}, {Sab{\'{\i}}n-Sanjuli{\'a}n}, {Smartt},
  {Stroud}, {van Loon}, {Vink}, \& {Walborn}}]{eva11}
{Evans}, C.~J., {Taylor}, W.~D., {H{\'e}nault-Brunet}, V., {et~al.} 2011, \aap,
  530, A108

\bibitem[{{Fossati} {et~al.}(2015{\natexlab{a}}){Fossati}, {Castro}, {Morel},
  {Langer}, {Briquet}, {Carroll}, {Hubrig}, {Nieva}, {Oskinova}, {Przybilla},
  {Schneider}, {Sch{\"o}ller}, {Sim{\'o}n-D{\'{\i}}az}, {Ilyin}, {de Koter},
  {Reisenegger}, \& {Sana}}]{fos15b}
{Fossati}, L., {Castro}, N., {Morel}, T., {et~al.} 2015{\natexlab{a}}, \aap,
  574, A20

\bibitem[{{Fossati} {et~al.}(2015{\natexlab{b}}){Fossati}, {Castro},
  {Sch{\"o}ller}, {Hubrig}, {Langer}, {Morel}, {Briquet}, {Herrero},
  {Przybilla}, {Sana}, {Schneider}, {de Koter}, \& {BOB
  Collaboration}}]{fos15a}
{Fossati}, L., {Castro}, N., {Sch{\"o}ller}, M., {et~al.} 2015{\natexlab{b}},
  \aap, 582, A45

\bibitem[{{Fossati} {et~al.}(2016){Fossati}, {Schneider}, {Castro}, {Langer},
  {Sim{\'o}n-D{\'{\i}}az}, {M{\"u}ller}, {de Koter}, {Morel}, {Petit}, {Sana},
  \& {Wade}}]{fos16}
{Fossati}, L., {Schneider}, F.~R.~N., {Castro}, N., {et~al.} 2016, \aap, 592,
  A84

\bibitem[{{Fraser} {et~al.}(2010){Fraser}, {Dufton}, {Hunter}, \&
  {Ryans}}]{fra10}
{Fraser}, M., {Dufton}, P.~L., {Hunter}, I., \& {Ryans}, R.~S.~I. 2010, \mnras,
  404, 1306

\bibitem[{{Frischknecht} {et~al.}(2010){Frischknecht}, {Hirschi}, {Meynet},
  {Ekstr{\"o}m}, {Georgy}, {Rauscher}, {Winteler}, \& {Thielemann}}]{fri10}
{Frischknecht}, U., {Hirschi}, R., {Meynet}, G., {et~al.} 2010, \aap, 522, A39

\bibitem[{{Garland} {et~al.}(2017){Garland}, {Dufton}, {Evans}, {Crowther},
  {Howarth}, {de Koter}, {de Mink}, {Grin}, {Langer}, {Lennon}, {McEvoy},
  {Sana}, {Schneider}, {S{\'{\i}}mon D{\'{\i}}az}, {Taylor}, {Thompson}, \&
  {Vink}}]{gar17}
{Garland}, R., {Dufton}, P.~L., {Evans}, C.~J., {et~al.} 2017, \aap, 603, A91

\bibitem[{{Garnett}(1999)}]{gar99}
{Garnett}, D.~R. 1999, in IAU Symposium, Vol. 190, New Views of the Magellanic
  Clouds, ed. {Y.-H.~Chu, N.~Suntzeff, J.~Hesser, \& D.~Bohlender}, 266

\bibitem[{{Georgy} {et~al.}(2013{\natexlab{a}}){Georgy}, {Ekstr{\"o}m},
  {Eggenberger}, {Meynet}, {Haemmerl{\'e}}, {Maeder}, {Granada}, {Groh},
  {Hirschi}, {Mowlavi}, {Yusof}, {Charbonnel}, {Decressin}, \&
  {Barblan}}]{geo13}
{Georgy}, C., {Ekstr{\"o}m}, S., {Eggenberger}, P., {et~al.}
  2013{\natexlab{a}}, \aap, 558, A103

\bibitem[{{Georgy} {et~al.}(2013{\natexlab{b}}){Georgy}, {Ekstr{\"o}m},
  {Granada}, {Meynet}, {Mowlavi}, {Eggenberger}, \& {Maeder}}]{geo13a}
{Georgy}, C., {Ekstr{\"o}m}, S., {Granada}, A., {et~al.} 2013{\natexlab{b}},
  \aap, 553, A24

\bibitem[{{Gray}(2005)}]{gra05}
{Gray}, D.~F. 2005, {The Observation and Analysis of Stellar Photospheres}

\bibitem[{{Gray}(2016)}]{gra16}
{Gray}, D.~F. 2016, \apj, 826, 92

\bibitem[{{Gray}(2017)}]{gra17}
{Gray}, D.~F. 2017, \apj, 845, 62

\bibitem[{{Grin} {et~al.}(2017){Grin}, {Ram{\'{\i}}rez-Agudelo}, {de Koter},
  {Sana}, {Puls}, {Brott}, {Crowther}, {Dufton}, {Evans}, {Gr{\"a}fener},
  {Herrero}, {Langer}, {Lennon}, {van Loon}, {Markova}, {de Mink}, {Najarro},
  {Schneider}, {Taylor}, {Tramper}, {Vink}, \& {Walborn}}]{gri17}
{Grin}, N.~J., {Ram{\'{\i}}rez-Agudelo}, O.~H., {de Koter}, A., {et~al.} 2017,
  \aap, 600, A82

\bibitem[{{Grunhut} {et~al.}(2017){Grunhut}, {Wade}, {Neiner}, {Oksala},
  {Petit}, {Alecian}, {Bohlender}, {Bouret}, {Henrichs}, {Hussain},
  {Kochukhov}, \& {MiMeS Collaboration}}]{gru17}
{Grunhut}, J.~H., {Wade}, G.~A., {Neiner}, C., {et~al.} 2017, \mnras, 465, 2432

\bibitem[{{Heger} \& {Langer}(2000)}]{heg00b}
{Heger}, A. \& {Langer}, N. 2000, \apj, 544, 1016

\bibitem[{{Henyey} {et~al.}(1959){Henyey}, {Lelevier}, \& {Levee}}]{hen59}
{Henyey}, L.~G., {Lelevier}, R., \& {Levee}, R.~D. 1959, \apj, 129, 2

\bibitem[{{Hirschi} {et~al.}(2004){Hirschi}, {Meynet}, \& {Maeder}}]{hir04}
{Hirschi}, R., {Meynet}, G., \& {Maeder}, A. 2004, \aap, 425, 649

\bibitem[{{Howarth} {et~al.}(1997){Howarth}, {Siebert}, {Hussain}, \&
  {Prinja}}]{how97}
{Howarth}, I.~D., {Siebert}, K.~W., {Hussain}, G.~A.~J., \& {Prinja}, R.~K.
  1997, \mnras, 284, 265

\bibitem[{{Hoyle}(1960)}]{hoy60}
{Hoyle}, F. 1960, \mnras, 120, 22

\bibitem[{{Huang} {et~al.}(2010){Huang}, {Gies}, \& {McSwain}}]{hua10}
{Huang}, W., {Gies}, D.~R., \& {McSwain}, M.~V. 2010, \apj, 722, 605

\bibitem[{{Hubeny}(1988)}]{hub88}
{Hubeny}, I. 1988, Computer Physics Communications, 52, 103

\bibitem[{{Hubeny} {et~al.}(1998){Hubeny}, {Heap}, \& {Lanz}}]{hub98}
{Hubeny}, I., {Heap}, S.~R., \& {Lanz}, T. 1998, in Astronomical Society of the
  Pacific Conference Series, Vol. 131, Properties of Hot Luminous Stars, ed.
  {I.~Howarth}, 108--+

\bibitem[{{Hubeny} \& {Lanz}(1995)}]{hub95}
{Hubeny}, I. \& {Lanz}, T. 1995, \apj, 439, 875

\bibitem[{{Hunter} {et~al.}(2009){Hunter}, {Brott}, {Langer}, {Lennon},
  {Dufton}, {Howarth}, {Ryans}, {Trundle}, {Evans}, {de Koter}, \&
  {Smartt}}]{hun09a}
{Hunter}, I., {Brott}, I., {Langer}, N., {et~al.} 2009, \aap, 496, 841

\bibitem[{{Hunter} {et~al.}(2008{\natexlab{a}}){Hunter}, {Brott}, {Lennon},
  {Langer}, {Dufton}, {Trundle}, {Smartt}, {de Koter}, {Evans}, \&
  {Ryans}}]{hun08b}
{Hunter}, I., {Brott}, I., {Lennon}, D.~J., {et~al.} 2008{\natexlab{a}}, \apjl,
  676, L29

\bibitem[{{Hunter} {et~al.}(2007){Hunter}, {Dufton}, {Smartt}, {Ryans},
  {Evans}, {Lennon}, {Trundle}, {Hubeny}, \& {Lanz}}]{hun07}
{Hunter}, I., {Dufton}, P.~L., {Smartt}, S.~J., {et~al.} 2007, \aap, 466, 277

\bibitem[{{Hunter} {et~al.}(2008{\natexlab{b}}){Hunter}, {Lennon}, {Dufton},
  {Trundle}, {Sim{\'o}n-D{\'{\i}}az}, {Smartt}, {Ryans}, \& {Evans}}]{hun08a}
{Hunter}, I., {Lennon}, D.~J., {Dufton}, P.~L., {et~al.} 2008{\natexlab{b}},
  \aap, 479, 541

\bibitem[{{Jackson} \& {Jeffries}(2010)}]{jac10}
{Jackson}, R.~J. \& {Jeffries}, R.~D. 2010, \mnras, 402, 1380

\bibitem[{{Korn} {et~al.}(2002){Korn}, {Keller}, {Kaufer}, {Langer},
  {Przybilla}, {Stahl}, \& {Wolf}}]{kor02}
{Korn}, A.~J., {Keller}, S.~C., {Kaufer}, A., {et~al.} 2002, \aap, 385, 143

\bibitem[{{Korn} {et~al.}(2005){Korn}, {Nieva}, {Daflon}, \& {Cunha}}]{kor05}
{Korn}, A.~J., {Nieva}, M.~F., {Daflon}, S., \& {Cunha}, K. 2005, \apj, 633,
  899

\bibitem[{{Korntreff} {et~al.}(2012){Korntreff}, {Kaczmarek}, \&
  {Pfalzner}}]{kor12}
{Korntreff}, C., {Kaczmarek}, T., \& {Pfalzner}, S. 2012, \aap, 543, A126

\bibitem[{{Kurt} \& {Dufour}(1998)}]{kur98}
{Kurt}, C.~M. \& {Dufour}, R.~J. 1998, in Revista Mexicana de Astronomia y
  Astrofisica Conference Series, Vol.~7, Revista Mexicana de Astronomia y
  Astrofisica Conference Series, ed. R.~J. {Dufour} \& S.~{Torres-Peimbert},
  202

\bibitem[{{Kushwaha}(1957)}]{kus57}
{Kushwaha}, R.~S. 1957, \apj, 125, 242

\bibitem[{{Langer}(2012)}]{lan12}
{Langer}, N. 2012, ARAA, 50, 107

\bibitem[{{Langer} {et~al.}(2008){Langer}, {Cantiello}, {Yoon}, {Hunter},
  {Brott}, {Lennon}, {de Mink}, \& {Verheijdt}}]{lan08}
{Langer}, N., {Cantiello}, M., {Yoon}, S.-C., {et~al.} 2008, in IAU Symposium,
  Vol. 250, IAU Symposium, ed. F.~{Bresolin}, P.~A. {Crowther}, \& J.~{Puls},
  167--178

\bibitem[{{Lanz} \& {Hubeny}(2003)}]{lan03}
{Lanz}, T. \& {Hubeny}, I. 2003, \apjs, 146, 417

\bibitem[{{Lanz} \& {Hubeny}(2007)}]{lan07}
{Lanz}, T. \& {Hubeny}, I. 2007, \apjs, 169, 83

\bibitem[{{Lemasle} {et~al.}(2017){Lemasle}, {Groenewegen}, {Grebel}, {Bono},
  {Fiorentino}, {Fran{\c c}ois}, {Inno}, {Kovtyukh}, {Matsunaga}, {Pedicelli},
  {Primas}, {Pritchard}, {Romaniello}, \& {da Silva}}]{lem17}
{Lemasle}, B., {Groenewegen}, M.~A.~T., {Grebel}, E.~K., {et~al.} 2017, ArXiv
  e-prints

\bibitem[{{Lennon} {et~al.}(2005){Lennon}, {Lee}, {Dufton}, \& {Ryans}}]{len05}
{Lennon}, D.~J., {Lee}, J.-K., {Dufton}, P.~L., \& {Ryans}, R.~S.~I. 2005,
  \aap, 438, 265

\bibitem[{{Lucy}(1974)}]{luc74}
{Lucy}, L.~B. 1974, \aj, 79, 745

\bibitem[{{Maeder}(1980)}]{mae80}
{Maeder}, A. 1980, \aap, 92, 101

\bibitem[{{Maeder}(1987)}]{mae87}
{Maeder}, A. 1987, \aap, 173, 247

\bibitem[{{Maeder}(2009)}]{mae09}
{Maeder}, A. 2009, {Physics, Formation and Evolution of Rotating Stars}

\bibitem[{{Maeder} {et~al.}(2012){Maeder}, {Georgy}, {Meynet}, \&
  {Ekstr{\"o}m}}]{mae12}
{Maeder}, A., {Georgy}, C., {Meynet}, G., \& {Ekstr{\"o}m}, S. 2012, \aap, 539,
  A110

\bibitem[{{Maeder} {et~al.}(2014){Maeder}, {Przybilla}, {Nieva}, {Georgy},
  {Meynet}, {Ekstr{\"o}m}, \& {Eggenberger}}]{mae14}
{Maeder}, A., {Przybilla}, N., {Nieva}, M.-F., {et~al.} 2014, \aap, 565, A39

\bibitem[{{Martayan} {et~al.}(2006){Martayan}, {Fr{\'e}mat}, {Hubert},
  {Floquet}, {Zorec}, \& {Neiner}}]{mar06}
{Martayan}, C., {Fr{\'e}mat}, Y., {Hubert}, A., {et~al.} 2006, \aap, 452, 273

\bibitem[{{Martayan} {et~al.}(2007){Martayan}, {Fr{\'e}mat}, {Hubert},
  {Floquet}, {Zorec}, \& {Neiner}}]{mart07}
{Martayan}, C., {Fr{\'e}mat}, Y., {Hubert}, A.-M., {et~al.} 2007, \aap, 462,
  683

\bibitem[{{McEvoy} {et~al.}(2015){McEvoy}, {Dufton}, {Evans}, {Kalari},
  {Markova}, {Sim{\'o}n-D{\'{\i}}az}, {Vink}, {Walborn}, {Crowther}, {de
  Koter}, {de Mink}, {Dunstall}, {H{\'e}nault-Brunet}, {Herrero}, {Langer},
  {Lennon}, {Ma{\'{\i}}z Apell{\'a}niz}, {Najarro}, {Puls}, {Sana},
  {Schneider}, \& {Taylor}}]{mce14}
{McEvoy}, C.~M., {Dufton}, P.~L., {Evans}, C.~J., {et~al.} 2015, \aap, 575, A70

\bibitem[{{Meynet} {et~al.}(2011){Meynet}, {Eggenberger}, \& {Maeder}}]{mey11}
{Meynet}, G., {Eggenberger}, P., \& {Maeder}, A. 2011, \aap, 525, L11

\bibitem[{{Meynet} \& {Maeder}(2000)}]{mey00}
{Meynet}, G. \& {Maeder}, A. 2000, \aap, 361, 101

\bibitem[{{Meynet} {et~al.}(2017){Meynet}, {Maeder}, {Georgy}, {Ekstr{\"o}m},
  {Eggenberger}, {Barblan}, \& {Song}}]{mey17}
{Meynet}, G., {Maeder}, A., {Georgy}, C., {et~al.} 2017, in IAU Symposium, Vol.
  329, The Lives and Death-Throes of Massive Stars, ed. J.~J. {Eldridge}, J.~C.
  {Bray}, L.~A.~S. {McClelland}, \& L.~{Xiao}, 3--14

\bibitem[{{Mokiem} {et~al.}(2007){Mokiem}, {de Koter}, {Vink}, {Puls}, {Evans},
  {Smartt}, {Crowther}, {Herrero}, {Langer}, {Lennon}, {Najarro}, \&
  {Villamariz}}]{mok07}
{Mokiem}, M.~R., {de Koter}, A., {Vink}, J.~S., {et~al.} 2007, \aap, 473, 603

\bibitem[{{Morel}(2012)}]{mor12}
{Morel}, T. 2012, in Astronomical Society of the Pacific Conference Series,
  Vol. 465, Proceedings of a Scientific Meeting in Honor of Anthony F. J.
  Moffat, ed. L.~{Drissen}, C.~{Robert}, N.~{St-Louis}, \& A.~F.~J. {Moffat},
  54

\bibitem[{{Morel} {et~al.}(2014){Morel}, {Castro}, {Fossati}, {Hubrig},
  {Langer}, {Przybilla}, {Sch{\"o}ller}, {Carroll}, {Ilyin}, {Irrgang},
  {Oskinova}, {Schneider}, {D{\'{\i}}az}, {Briquet}, {Gonz{\'a}lez},
  {Kharchenko}, {Nieva}, {Scholz}, {de Koter}, {Hamann}, {Herrero},
  {Ma{\'{\i}}z Apell{\'a}niz}, {Sana}, {Arlt}, {Barb{\'a}}, {Dufton},
  {Kholtygin}, {Mathys}, {Piskunov}, {Reisenegger}, {Spruit}, \&
  {Yoon}}]{mor14}
{Morel}, T., {Castro}, N., {Fossati}, L., {et~al.} 2014, The Messenger, 157, 27

\bibitem[{{Morel} {et~al.}(2015){Morel}, {Castro}, {Fossati}, {Hubrig},
  {Langer}, {Przybilla}, {Sch{\"o}ller}, {Carroll}, {Ilyin}, {Irrgang},
  {Oskinova}, {Schneider}, {D{\'{\i}}az}, {Briquet}, {Gonz{\'a}lez},
  {Kharchenko}, {Nieva}, {Scholz}, {de Koter}, {Hamann}, {Herrero},
  {Ma{\'{\i}}z Apell{\'a}niz}, {Sana}, {Arlt}, {Barb{\'a}}, {Dufton},
  {Kholtygin}, {Mathys}, {Piskunov}, {Reisenegger}, {Spruit}, \&
  {Yoon}}]{mor15}
{Morel}, T., {Castro}, N., {Fossati}, L., {et~al.} 2015, in IAU Symposium, Vol.
  307, New Windows on Massive Stars, ed. G.~{Meynet}, C.~{Georgy}, J.~{Groh},
  \& P.~{Stee}, 342--347

\bibitem[{{Morel} {et~al.}(2008){Morel}, {Hubrig}, \& {Briquet}}]{mor08}
{Morel}, T., {Hubrig}, S., \& {Briquet}, M. 2008, \aap, 481, 453

\bibitem[{{Nieva} \& {Przybilla}(2006)}]{nie06}
{Nieva}, M.~F. \& {Przybilla}, N. 2006, \apjl, 639, L39

\bibitem[{{Palma} {et~al.}(2015){Palma}, {Clari{\'a}}, {Geisler}, {Gramajo}, \&
  {Ahumada}}]{pal15}
{Palma}, T., {Clari{\'a}}, J.~J., {Geisler}, D., {Gramajo}, L.~V., \&
  {Ahumada}, A.~V. 2015, \mnras, 450, 2122

\bibitem[{{Pasquini} {et~al.}(2002){Pasquini}, {Avila}, {Blecha}, {Cacciari},
  {Cayatte}, {Colless}, {Damiani}, {de Propris}, {Dekker}, {di Marcantonio},
  {Farrell}, {Gillingham}, {Guinouard}, {Hammer}, {Kaufer}, {Hill}, {Marteaud},
  {Modigliani}, {Mulas}, {North}, {Popovic}, {Rossetti}, {Royer}, {Santin},
  {Schmutzer}, {Simond}, {Vola}, {Waller}, \& {Zoccali}}]{pas02}
{Pasquini}, L., {Avila}, G., {Blecha}, A., {et~al.} 2002, The Messenger, 110, 1

\bibitem[{{Penny}(1996)}]{pen96}
{Penny}, L.~R. 1996, \apj, 463, 737

\bibitem[{{Petermann} {et~al.}(2015){Petermann}, {Langer}, {Castro}, \&
  {Fossati}}]{pet15}
{Petermann}, I., {Langer}, N., {Castro}, N., \& {Fossati}, L. 2015, \aap, 584,
  A54

\bibitem[{{Pietrzy{\'n}ski} {et~al.}(2013){Pietrzy{\'n}ski}, {Graczyk},
  {Gieren}, {Thompson}, {Pilecki}, {Udalski}, {Soszy{\'n}ski}, {Koz{\l}owski},
  {Konorski}, {Suchomska}, {Bono}, {Moroni}, {Villanova}, {Nardetto},
  {Bresolin}, {Kudritzki}, {Storm}, {Gallenne}, {Smolec}, {Minniti}, {Kubiak},
  {Szyma{\'n}ski}, {Poleski}, {Wyrzykowski}, {Ulaczyk}, {Pietrukowicz},
  {G{\'o}rski}, \& {Karczmarek}}]{pie13}
{Pietrzy{\'n}ski}, G., {Graczyk}, D., {Gieren}, W., {et~al.} 2013, \nat, 495,
  76

\bibitem[{{Podsiadlowski} {et~al.}(1992){Podsiadlowski}, {Joss}, \&
  {Hsu}}]{pod92}
{Podsiadlowski}, P., {Joss}, P.~C., \& {Hsu}, J.~J.~L. 1992, \apj, 391, 246

\bibitem[{{Potter} {et~al.}(2012{\natexlab{a}}){Potter}, {Chitre}, \&
  {Tout}}]{pot12b}
{Potter}, A.~T., {Chitre}, S.~M., \& {Tout}, C.~A. 2012{\natexlab{a}}, \mnras,
  424, 2358

\bibitem[{{Potter} {et~al.}(2012{\natexlab{b}}){Potter}, {Tout}, \&
  {Brott}}]{pot12a}
{Potter}, A.~T., {Tout}, C.~A., \& {Brott}, I. 2012{\natexlab{b}}, \mnras, 423,
  1221

\bibitem[{{Przybilla} \& {Nieva}(2011)}]{prz11}
{Przybilla}, N. \& {Nieva}, M.-F. 2011, in IAU Symposium, Vol. 272, Active OB
  Stars: Structure, Evolution, Mass Loss, and Critical Limits, ed. C.~{Neiner},
  G.~{Wade}, G.~{Meynet}, \& G.~{Peters}, 26--31

\bibitem[{{Puls} {et~al.}(2008){Puls}, {Vink}, \& {Najarro}}]{pul08}
{Puls}, J., {Vink}, J.~S., \& {Najarro}, F. 2008, \aapr, 16, 209

\bibitem[{{Ram{\'{\i}}rez-Agudelo} {et~al.}(2015){Ram{\'{\i}}rez-Agudelo},
  {Sana}, {de Mink}, {H{\'e}nault-Brunet}, {de Koter}, {Langer}, {Tramper},
  {Gr{\"a}fener}, {Evans}, {Vink}, {Dufton}, \& {Taylor}}]{ram15}
{Ram{\'{\i}}rez-Agudelo}, O.~H., {Sana}, H., {de Mink}, S.~E., {et~al.} 2015,
  \aap, 580, A92

\bibitem[{{Ram{\'{\i}}rez-Agudelo} {et~al.}(2013){Ram{\'{\i}}rez-Agudelo},
  {Sim{\'o}n-D{\'{\i}}az}, {Sana}, {de Koter}, {Sab{\'{\i}}n-Sanjul{\'{\i}}an},
  {de Mink}, {Dufton}, {Gr{\"a}fener}, {Evans}, {Herrero}, {Langer}, {Lennon},
  {Ma{\'{\i}}z Apell{\'a}niz}, {Markova}, {Najarro}, {Puls}, {Taylor}, \&
  {Vink}}]{ram13}
{Ram{\'{\i}}rez-Agudelo}, O.~H., {Sim{\'o}n-D{\'{\i}}az}, S., {Sana}, H.,
  {et~al.} 2013, \aap, 560, A29

\bibitem[{{Rivero Gonz{\'a}lez} {et~al.}(2012){Rivero Gonz{\'a}lez}, {Puls},
  {Najarro}, \& {Brott}}]{riv12}
{Rivero Gonz{\'a}lez}, J.~G., {Puls}, J., {Najarro}, F., \& {Brott}, I. 2012,
  \aap, 537, A79

\bibitem[{{Ryans} {et~al.}(2003){Ryans}, {Dufton}, {Mooney}, {Rolleston},
  {Keenan}, {Hubeny}, \& {Lanz}}]{rya03}
{Ryans}, R.~S.~I., {Dufton}, P.~L., {Mooney}, C.~J., {et~al.} 2003, \aap, 401,
  1119

\bibitem[{{Salpeter}(1955)}]{sal55}
{Salpeter}, E.~E. 1955, \apj, 121, 161

\bibitem[{{Schneider} {et~al.}(2014){Schneider}, {Langer}, {de Koter}, {Brott},
  {Izzard}, \& {Lau}}]{sch14}
{Schneider}, F.~R.~N., {Langer}, N., {de Koter}, A., {et~al.} 2014, \aap, 570,
  A66

\bibitem[{{Schneider} {et~al.}(2016){Schneider}, {Podsiadlowski}, {Langer},
  {Castro}, \& {Fossati}}]{sch16}
{Schneider}, F.~R.~N., {Podsiadlowski}, P., {Langer}, N., {Castro}, N., \&
  {Fossati}, L. 2016, \mnras, 457, 2355

\bibitem[{{Sch{\"o}ller} {et~al.}(2017){Sch{\"o}ller}, {Hubrig}, {Fossati},
  {Carroll}, {Briquet}, {Oskinova}, {J{\"a}rvinen}, {Ilyin}, {Castro}, {Morel},
  {Langer}, {Przybilla}, {Nieva}, {Kholtygin}, {Sana}, {Herrero}, {Barb{\'a}},
  {de Koter}, \& {BOB Collaboration}}]{sch17}
{Sch{\"o}ller}, M., {Hubrig}, S., {Fossati}, L., {et~al.} 2017, \aap, 599, A66

\bibitem[{{Schwarzschild} \& {H{\"a}rm}(1958)}]{sch58}
{Schwarzschild}, M. \& {H{\"a}rm}, R. 1958, \apj, 128, 348

\bibitem[{{Sigut}(1996)}]{sig96}
{Sigut}, T.~A.~A. 1996, \apj, 473, 452

\bibitem[{{Sim{\'o}n-D{\'{\i}}az}
  {et~al.}(2017{\natexlab{a}}){Sim{\'o}n-D{\'{\i}}az}, {Aerts}, {Urbaneja},
  {Camacho}, {Antoci}, {Fredslund Andersen}, {Grundahl}, \&
  {Pall{\'e}}}]{sim17a}
{Sim{\'o}n-D{\'{\i}}az}, S., {Aerts}, C., {Urbaneja}, M.~A., {et~al.}
  2017{\natexlab{a}}, ArXiv e-prints

\bibitem[{{Sim{\'o}n-D{\'{\i}}az}
  {et~al.}(2017{\natexlab{b}}){Sim{\'o}n-D{\'{\i}}az}, {Godart}, {Castro},
  {Herrero}, {Aerts}, {Puls}, {Telting}, \& {Grassitelli}}]{sim17}
{Sim{\'o}n-D{\'{\i}}az}, S., {Godart}, M., {Castro}, N., {et~al.}
  2017{\natexlab{b}}, \aap, 597, A22

\bibitem[{{Sim{\'o}n-D{\'{\i}}az} \& {Herrero}(2007)}]{sim07}
{Sim{\'o}n-D{\'{\i}}az}, S. \& {Herrero}, A. 2007, \aap, 468, 1063

\bibitem[{{Sim{\'o}n-D{\'{\i}}az} \& {Herrero}(2014)}]{sim14}
{Sim{\'o}n-D{\'{\i}}az}, S. \& {Herrero}, A. 2014, \aap, 562, A135

\bibitem[{{Sim{\'o}n-D{\'{\i}}az} {et~al.}(2010){Sim{\'o}n-D{\'{\i}}az},
  {Herrero}, {Uytterhoeven}, {Castro}, {Aerts}, \& {Puls}}]{sim10}
{Sim{\'o}n-D{\'{\i}}az}, S., {Herrero}, A., {Uytterhoeven}, K., {et~al.} 2010,
  \apjl, 720, L174

\bibitem[{{Slettebak}(1949)}]{sle49}
{Slettebak}, A. 1949, \apj, 110, 498

\bibitem[{{Spruit}(2002)}]{spr02}
{Spruit}, H.~C. 2002, \aap, 381, 923

\bibitem[{{Stahler}(2010)}]{sta10}
{Stahler}, S.~W. 2010, \mnras, 402, 1758

\bibitem[{{Struve}(1931)}]{str31}
{Struve}, O. 1931, \apj, 73, 94

\bibitem[{{Sz{\'e}csi} {et~al.}(2015){Sz{\'e}csi}, {Langer}, {Yoon}, {Sanyal},
  {de Mink}, {Evans}, \& {Dermine}}]{sze15}
{Sz{\'e}csi}, D., {Langer}, N., {Yoon}, S.-C., {et~al.} 2015, \aap, 581, A15

\bibitem[{{Tayler}(1954)}]{tay54}
{Tayler}, R.~J. 1954, \apj, 120, 332

\bibitem[{{Tayler}(1956)}]{tay56}
{Tayler}, R.~J. 1956, \mnras, 116, 25

\bibitem[{{Taylor} {et~al.}(2014){Taylor}, {Evans}, {Sim{\'o}n-D{\'{\i}}az},
  {Sana}, {Langer}, {Smith}, \& {Smartt}}]{tay14}
{Taylor}, W.~D., {Evans}, C.~J., {Sim{\'o}n-D{\'{\i}}az}, S., {et~al.} 2014,
  \mnras, 442, 1483

\bibitem[{{Townsend} {et~al.}(2004){Townsend}, {Owocki}, \& {Howarth}}]{tow04}
{Townsend}, R.~H.~D., {Owocki}, S.~P., \& {Howarth}, I.~D. 2004, \mnras, 350,
  189

\bibitem[{{Trundle} {et~al.}(2007{\natexlab{a}}){Trundle}, {Dufton}, {Hunter},
  {Evans}, {Lennon}, {Smartt}, \& {Ryans}}]{tru08}
{Trundle}, C., {Dufton}, P.~L., {Hunter}, I., {et~al.} 2007{\natexlab{a}},
  \aap, 471, 625

\bibitem[{{Trundle} {et~al.}(2007{\natexlab{b}}){Trundle}, {Dufton}, {Hunter},
  {Evans}, {Lennon}, {Smartt}, \& {Ryans}}]{tru07}
{Trundle}, C., {Dufton}, P.~L., {Hunter}, I., {et~al.} 2007{\natexlab{b}},
  \aap, 471, 625

\bibitem[{{Ud-Doula} {et~al.}(2009){Ud-Doula}, {Owocki}, \& {Townsend}}]{udd09}
{Ud-Doula}, A., {Owocki}, S.~P., \& {Townsend}, R.~H.~D. 2009, \mnras, 392,
  1022

\bibitem[{{Wade} {et~al.}(2014){Wade}, {Grunhut}, {Alecian}, {Neiner},
  {Auri{\`e}re}, {Bohlender}, {David-Uraz}, {Folsom}, {Henrichs}, {Kochukhov},
  {Mathis}, {Owocki}, {Petit}, \& {Petit}}]{wad14}
{Wade}, G.~A., {Grunhut}, J., {Alecian}, E., {et~al.} 2014, in IAU Symposium,
  Vol. 302, Magnetic Fields throughout Stellar Evolution, ed. P.~{Petit},
  M.~{Jardine}, \& H.~C. {Spruit}, 265--269

\bibitem[{{Wade} {et~al.}(2016){Wade}, {Neiner}, {Alecian}, {Grunhut}, {Petit},
  {de Batz}, {Bohlender}, {Cohen}, {Henrichs}, {Kochukhov}, {Landstreet},
  {Manset}, {Martins}, {Mathis}, {Oksala}, {Owocki}, {Rivinius}, {Shultz},
  {Sundqvist}, {Townsend}, {ud-Doula}, {Bouret}, {Braithwaite}, {Briquet},
  {Carciofi}, {David-Uraz}, {Folsom}, {Fullerton}, {Leroy}, {Marcolino},
  {Moffat}, {Naz{\'e}}, {Louis}, {Auri{\`e}re}, {Bagnulo}, {Bailey},
  {Barb{\'a}}, {Blaz{\`e}re}, {B{\"o}hm}, {Catala}, {Donati}, {Ferrario},
  {Harrington}, {Howarth}, {Ignace}, {Kaper}, {L{\"u}ftinger}, {Prinja},
  {Vink}, {Weiss}, \& {Yakunin}}]{wad16}
{Wade}, G.~A., {Neiner}, C., {Alecian}, E., {et~al.} 2016, \mnras, 456, 2

\bibitem[{{Walborn} \& {Blades}(1997)}]{wal97}
{Walborn}, N.~R. \& {Blades}, J.~C. 1997, \apjs, 112, 457

\bibitem[{{Wegner}(1994)}]{weg94}
{Wegner}, W. 1994, \mnras, 270, 229

\bibitem[{{Woosley} \& {Heger}(2006)}]{woo06}
{Woosley}, S.~E. \& {Heger}, A. 2006, \apj, 637, 914

\bibitem[{{Yoon} \& {Langer}(2005)}]{yoo05}
{Yoon}, S.-C. \& {Langer}, N. 2005, \aap, 443, 643

\bibitem[{{Yoon} {et~al.}(2006){Yoon}, {Langer}, \& {Norman}}]{yoo06}
{Yoon}, S.-C., {Langer}, N., \& {Norman}, C. 2006, \aap, 460, 199

\end{thebibliography}

\end{document}